\documentclass{ieeeaccess}
\usepackage{cite}
\usepackage{amsmath,amssymb,amsfonts}
\usepackage{graphicx}
\usepackage{textcomp}
\def\BibTeX{{\rm B\kern-.05em{\sc i\kern-.025em b}\kern-.08em
    T\kern-.1667em\lower.7ex\hbox{E}\kern-.125emX}}

\usepackage{booktabs} 
\usepackage{adjustbox}
\usepackage{multirow}
\usepackage{hyperref}       
\usepackage{url}            
\usepackage{nicefrac}       
\usepackage{microtype}      
\usepackage{xcolor}         
\usepackage[font=footnotesize,skip=1pt]{caption}
\usepackage[ruled,vlined]{algorithm2e}
\usepackage{comment}
\usepackage{braket}
\usepackage{float}
\begin{document}
\pagestyle{empty}

\history{~}
\doi{~}

\title{Q-PhotoNAS: Hybrid Quantum Neural Architecture Search Framework on Photonic Devices}

\author{\uppercase{Farah Elnakhal}\authorrefmark{1,2}
, \uppercase{Alberto Marchisio}\authorrefmark{2,3}\IEEEmembership{Member, IEEE}
, \uppercase{Nouhaila Innan}\authorrefmark{2,3}\IEEEmembership{Member, IEEE}
, \uppercase{Gabriel Falcao}\authorrefmark{1,4}\IEEEmembership{Senior Member, IEEE}
, \uppercase{Muhammad Shafique}\authorrefmark{2,3}\IEEEmembership{Senior Member, IEEE}
}

\address[1]{Science Division, New York University Abu Dhabi, UAE}
\address[2]{eBrain Lab, Division of Engineering, New York University Abu Dhabi, PO Box 129188, Abu Dhabi, UAE}
\address[3]{Center for Quantum and Topological Systems, NYUAD Research
Institute, New York University Abu Dhabi, UAE}
\address[4]{Instituto de Telecomunicações, University of Coimbra, Portugal}

\tfootnote{This work was supported in part by the NYUAD Center for Interdisciplinary Data Science \& AI (CIDSAI), funded by Tamkeen under the NYUAD Research Institute Award CG016, and by the NYUAD Center for Quantum and Topological Systems (CQTS), funded by Tamkeen under the NYUAD Research Institute grant CG008. This work is partially funded by national funds through FCT – Fundação para a Ciência e a Tecnologia, I.P., and, when eligible, co-funded by EU funds under project/support UID/50008/2025 – Instituto de Telecomunicações, with DOI identifier - https://doi.org/10.54499/UID/50008/2025.}

\markboth
{F. Elnakhal \headeretal: Q-PhotoNAS: Hybrid Quantum Neural Architecture Search Framework on Photonic Devices}
{F. Elnakhal \headeretal: Q-PhotoNAS: Hybrid Quantum Neural Architecture Search Framework on Photonic Devices}

\corresp{Corresponding author: Alberto Marchisio (email: alberto.marchisio@nyu.edu).}

\begin{abstract}
Photonic quantum computing is a promising platform for scalable quantum machine learning, but designing effective hybrid architectures remains challenging under hardware and optimization constraints. Existing approaches rely on manually tuned architectures that fail to account for the collaboration between classical preprocessing, phase encoding, and photonic circuit structure, limiting both accuracy and hardware compatibility. In this paper, we propose a neural architecture search framework for hybrid photonic quantum-classical models that combines genetic algorithm-based search with learnable quantum phase encoding to systematically explore the joint design space of classical and quantum components. Our framework encodes 19 hyperparameters across six gene groups and evolves a population of hybrid architectures using group-based crossover, per-gene mutation, and elitism, evaluating each candidate on a short training budget before full retraining of the best found design. We evaluate our framework on two image classification benchmarks, Digits and MNIST, achieving final validation accuracies of 99.44\% and 98.78\%, respectively, with first-principles execution time estimates on the Quandela Ascella photonic QPU projecting single-image inference at $\sim$ 67 ms (Digits) and $\sim$ 149 ms (MNIST). Our quantum contribution analysis further shows that the photonic layer extracts non-redundant features orthogonal to the classical pathway, providing a measurable accuracy advantage over classical-only baselines. Our results demonstrate that automated architecture search is both practical and impactful for hybrid photonic systems, opening the way for systematic design space exploration of quantum AI on photonic devices.

\end{abstract}



\maketitle
\thispagestyle{empty}

\section{Introduction}
\label{sec:introduction}

\PARstart{Q}{uantum} machine learning (QML) has emerged as one of the most actively explored intersections of two transformative technologies: quantum computing and artificial intelligence~\cite{biamonte2017quantum,cerezo2022challenges,zaman2023survey}. By offloading parts of a learning pipeline onto quantum hardware, QML aims to utilize phenomena such as superposition, interference, and entanglement to extract features that are fundamentally inaccessible to classical circuits of comparable size~\cite{schuld2021machine}. Quantum computation can be realized across several hardware platforms, including superconducting circuits, trapped ions, neutral atoms, spin systems, and photonic processors, each implementing quantum gates through different physical mechanisms~\cite{kjaergaard2020superconducting,bruzewicz2019trapped,flamini2019photonic,slussarenko2019photonic}. Among the physical platforms being developed for this purpose, photonic quantum computing is particularly relevant because it operates at room temperature, suffers from low decoherence compared with superconducting or trapped ion systems, and is compatible with both classical fiber networks and quantum communication infrastructure~\cite{zhangPhotonic2024,maring2023ascella}. 
Photonic processors encode quantum information in the states of photons and manipulate it using linear optical components such as beam splitters and phase shifters, and using measurement-based operations, which makes them naturally resilient to decoherence~\cite{flamini2019photonic,slussarenko2019photonic}. These properties have motivated a growing body of work demonstrating QML algorithms on photonic devices, ranging from kernel methods to variational quantum classifiers~\cite{proxiML2024,qorc2025}.

Despite these promising demonstrations, \textbf{achieving competitive accuracy on real photonic hardware remains difficult.} The core challenge is that a hybrid photonic quantum-classical model involves a large number of dependent design choices spanning: (i)~the classical preprocessing pipeline that reduces raw data to a form compatible with the photonic layer's hard input-size constraints; (ii)~the data encoding strategy that maps classical features to optical phase values that feed the photonic phase shifters; and (iii)~the structure of classical layers surrounding the quantum circuit, including depth, width, activation functions, normalization, dropout, and training hyperparameters. These choices interact strongly, and the consequences of getting them wrong are severe.

\begin{figure}[t!]
    \centering
    \includegraphics[width=\columnwidth]{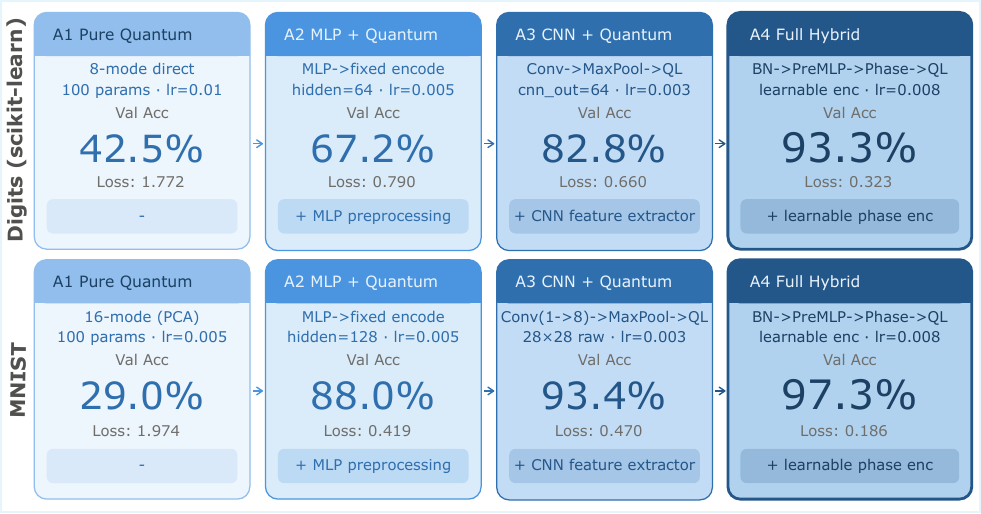}
    \caption{Motivation for automated architecture search in hybrid photonic QML. Both Digits (top) and MNIST (bottom) require four successive hand-designed attempts before reaching competitive accuracy. Each attempt is architecturally distinct from the last and requires its own training run.}
    \label{fig:prog_dev_motivation}
\end{figure}

Fig.~\ref{fig:prog_dev_motivation} tells this story directly. On Digits (top), a pure quantum circuit with no classical wrapper achieves only 42.5\% accuracy. Adding MLP preprocessing raises this to 67.2\%, a CNN feature extractor pushes further to 82.8\%, and only the full hybrid stack with learnable phase encoding reaches a useful 93.3\%. Each step is structurally distinct from the last and requires its own training run. On MNIST (bottom), the same pattern holds: a pure quantum circuit reaches only 29.0\%, MLP preprocessing jumps this to 88.0\%, a CNN extractor adds a further gain to 93.4\%, and the full hybrid model with learnable encoding culminates at 97.3\%. \textbf{Each step took time, each attempt was structurally different from the last, and there was no systematic way to know in advance which changes would help.} This is precisely where automated search is needed: the design space is too large for exhaustive enumeration (${\approx}3.7\times10^{10}$ configurations), the interaction between components makes greedy tuning unreliable, and the cost of each wrong attempt is a full training run. Manual tuning is therefore not a viable design strategy for hybrid photonic systems.

These challenges of joint component optimization and fixed non-adaptive encoding motivate this paper. While neural architecture search (NAS) has revolutionized classical deep learning~\cite{white2023nas,wang2024nasadvances}, and preliminary NAS frameworks have been proposed for gate-based quantum circuits~\cite{nasqcircuit2023,bqnas2024,dutta2025qas,choudhary2025graph,kashif2025faqnas,ahmed2026gat}, \textbf{no existing automated tools specifically target photonic hardware design space exploration}, where the
relevant design axes (mode count, encoding nonlinearity, pre-quantum MLP depth) differ fundamentally from those of qubit-based systems.
  
To address the above challenges, \textbf{we propose a Neural Architecture Search framework for hybrid Photonic quantum-classical Networks}, which makes the following key contributions:
 
\begin{enumerate}
    \item \textbf{Genetic algorithm-based joint design space search (Section~\ref{sec:framework})}: We encode 19 hyperparameters across six functional gene groups spanning classical preprocessing, phase encoding, photonic layer, and classifier head, and evolve a population of 20 candidate architectures over 30 generations using group-based crossover, per-gene mutation, and elitism, over a search space of approximately $3.7 \times 10^{10}$ configurations.
 
    \item \textbf{Learnable quantum phase encoding as a jointly optimized design axis (Section~\ref{subsec:Learnable Quantum  Phase Encoding})}: We replace fixed encoding schemes with a differentiable, per-feature scale-and-bias transformation whose activation function, scale initialization, and bias inclusion are all searched by the GA end-to-end alongside all other architectural choices. This joint optimization enables the photonic phase shifters to receive phases that are simultaneously compatible with the hardware range and aligned with the learned data distribution.
 
    \item \textbf{Hardware execution time estimation on Quandela Ascella QPU (Section~\ref{sec:results})}: We provide first-principles and Monte Carlo timing estimates for the optimized architectures on the Quandela Ascella photonic QPU~\cite{maring2023ascella}, confirming single-image inference in the tens-of-milliseconds regime and identifying thermal phase-shifter reconfiguration as the dominant hardware cost.
\end{enumerate}
 
We evaluate our framework on two image classification benchmarks: Scikit-learn Digits dataset and MNIST, the latter being a standard benchmark in both classical and quantum machine learning due to its widespread adoption for comparing hybrid quantum-classical models. We also provide quantum contribution analysis showing that the photonic layer delivers a measurable accuracy advantage over a parameter-matched classical-only baseline.
\textbf{To the best of our knowledge, this is the first NAS framework specifically designed for hybrid photonic quantum-classical architectures}, enabling systematic design space exploration on quantum hardware.
 
\section{Background and Related Work}
\label{sec:background}

\subsection{Photonic Quantum Computing and Machine Learning}

\begin{figure}[t!]
    \centering
    \includegraphics[width=\columnwidth]{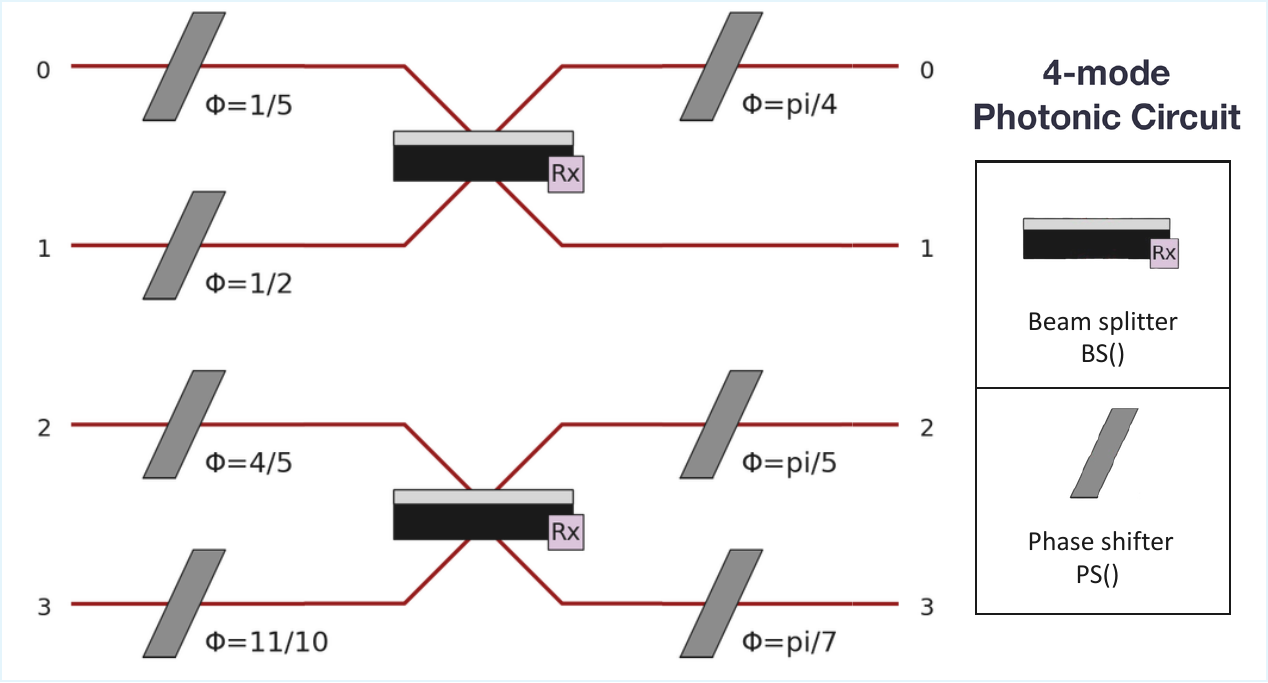}
    \caption{Simple 4-mode photonic circuit. PS($\phi$) applies a phase shift $\phi$ to a single mode. BS() is a balanced beam splitter that couples two modes. The circuit applies: PS(0.2,0.5,0.8,1.1) (phase values in radians) to modes 0–3; BS on pairs (0,1) and (2,3); PS($\pi/4$, $\pi/5$) to modes 0,2; BS on (2,3); then PS($\pi/6$, $\pi/7$) to modes 1,3. This sequence implements a programmable unitary transformation for photonic QML.}
    \label{fig:photonic_circuit}
\end{figure}

Photonic quantum computers manipulate information encoded in the quantum states of photons, exploiting linear-optical elements (beam splitters, phase shifters, and single-photon detectors) to implement computations that are classically difficult to simulate~\cite{zhangPhotonic2024}. The Quandela Ascella processor~\cite{maring2023ascella}, which we target in this work, is a cloud-accessible 12-mode silicon nitride chip equipped with 126 thermo-optic phase shifters, 132 directional couplers, and high-efficiency superconducting nanowire single-photon detectors; it supports both gate-based and photon-native (boson-sampling-style) computation. This distribution is believed to be classically hard to simulate, forming the basis of photonic quantum advantage claims~\cite{zhangPhotonic2024}. Fig. ~\ref{fig:photonic_circuit} shows a representative 4-mode programmable circuit with phase shifters (PS) and beam splitters (BS).

Early QML work on photonic hardware focused on boson-sampling-based feature extraction~\cite{qorc2025} and kernel methods~\cite{yin2025kernel}, which exploit multi-photon interference to construct quantum feature maps that are believed to be hard to reproduce classically. ProxiML~\cite{proxiML2024} demonstrated over 90\% accuracy on four-class classification tasks on Xanadu's X8 photonic device by introducing noise-aware design elements; however, its architecture was hand-crafted and not jointly optimized with the classical wrapper. The Perceval Challenge~\cite{perceval2025} established the first open benchmark for photonic QML on MNIST-like tasks, cataloging variational, hardware-native, and hybrid approaches and revealing that no single architecture dominates across all metrics, motivating a systematic search over the joint design space. More recently, Liu et al.~\cite{liu2024qadvantage} demonstrated a provable photonic quantum learning advantage using entangled probes, reducing sample complexity by more than 11 orders of magnitude compared with classical methods, while Yin et al.~\cite{yin2025kernel} showed experimentally that a photonic kernel classifier can outperform Gaussian and neural tangent kernels on binary classification tasks.
These results collectively establish the practical relevance of photonic QML but leave open the question of how to systematically design the hybrid architectures that wrap the quantum layer.
\subsection{Hybrid Quantum-Classical Neural Networks}

\begin{figure}[t!]
    \centering
    \includegraphics[width=\columnwidth]{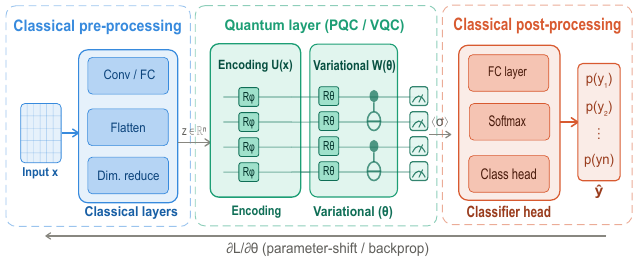}
    \caption{Generic hybrid quantum-classical neural network. Input $x$ passes through classical layers (Conv/FC) to produce latent vector $z \in \mathbb{R}^n$. Encoding circuit $U(x)$ maps $z$ to a quantum state. Variational circuit $W(\theta)$ with trainable parameters $\theta$ applies a unitary. Measurement yields expectation values $\langle \sigma \rangle$ (observable averages). A classical head (FC + Softmax) outputs probabilities $p(y_i)$ and prediction $\hat{y}$. Gradients $\partial L/\partial \theta$ are computed via parameter-shift rule, enabling standard backpropagation through the quantum layer without requiring knowledge of the circuit's internal structure. This is a general representation of HQCNNs.}
    \label{fig:hqcnn_architecture}
\end{figure}

Hybrid quantum-classical neural networks (HQCNNs) embed a parameterized quantum circuit (PQC) or variational quantum circuit (VQC) as a differentiable layer within an otherwise classical neural network, allowing end-to-end gradient-based training~\cite{cerezo2022challenges,senokosov2024encoding, Monbroussou_2025,innan2025next,vyskubov2026scaling}. Classical layers before the quantum circuit preprocess and dimensionally reduce the input, while layers after it map quantum measurement outcomes to class predictions. Existing work on HQCNNs for image classification has predominantly targeted gate-based (qubit) hardware: shallow HQCNN models have been proposed for MNIST~\cite{wang2024shallowHQ}, FashionMNIST~\cite{noiseHQCNN2024}, MedMNIST~\cite{qcqcnn2025}, and radiological CT imaging~\cite{matic2022qccnn}, and lean four-layer VQC architectures have achieved 100\% accuracy on MNIST with significantly fewer parameters than classical counterparts~\cite{liu2025lcqhnn}. On gate-based platforms, the choice of data encoding (angle, amplitude, or basis) has been shown to affect classification accuracy and qubit requirements~\cite{senokosov2024encoding,bose2025qdata,kashif2026design}, motivating learnable encoding schemes. Figure~\ref{fig:hqcnn_architecture} illustrates this flow, where classical preprocessing, quantum encoding ($U(x)$), variational processing ($W(\theta)$), measurement ($\langle \sigma \rangle$), and a classical classifier head are chained end-to-end.

Photonic HQCNNs present additional design challenges because photonic layers impose hard input-size constraints (the number of modes), produce photon-number probability vectors rather than expectation values, and are subject to photon loss and detection inefficiency. The Perceval Challenge~\cite{perceval2025} explored a range of hybrid strategies on a reduced MNIST task and showed that post-quantum classical processing contributes significantly to overall accuracy.
 
\subsection{Neural Architecture Search}
 
Neural architecture search (NAS) automates the design of neural network architectures by framing architecture selection as an optimization problem over a discrete or continuous search space~\cite{white2023nas,wang2024nasadvances}. The three canonical NAS strategies are reinforcement learning (RL), differentiable search (e.g., DARTS), and evolutionary/genetic algorithms (EA-NAS). Evolutionary NAS methods encode candidate architectures as genomes and apply biological operators (selection, crossover, and mutation) to explore the search space, offering robustness to non-differentiable objectives and discrete hyperparameters~\cite{vincent2023automl,L2024evolving}. GA-based hyperparameter optimization has been applied to CNN architecture search~\cite{cho2021gadl}, achieving performance competitive with expert-designed networks on MNIST and medical imaging benchmarks. Short proxy-budget training is a widely used and practical approximation~\cite{white2023nas}.

In the quantum domain, hierarchical NAS representations for parameterized quantum circuits have been proposed~\cite{nasqcircuit2023}, employing genetic algorithms to perform Quantum Phase Recognition and showing that circuit architecture has a significant impact on QML accuracy. Quantum-inspired evolutionary NAS (Q-NAS)~\cite{qnas2020} demonstrated that quantum probability amplitudes can be leveraged to accelerate architecture search, reaching 93.85\% on CIFAR-10. A more recent balanced quantum NAS approach (BQNAS)~\cite{bqnas2024} introduced alternate supernet training to mitigate weight-sharing inconsistencies. However, none of these quantum NAS approaches target photonic hardware or the joint design space of classical preprocessing, learnable phase encoding, and photonic circuit structure, which is the focus of this work. Our framework fills this gap by designing the genome, crossover operators, and fitness evaluation protocol specifically for hybrid photonic quantum-classical architectures, where group-based crossover is essential to preserve parameters within each architectural component.
 
\section{Q-PhotoNAS Framework}
\label{sec:framework}
\begin{figure}[t!]
    \centering
    \includegraphics[width=\columnwidth]{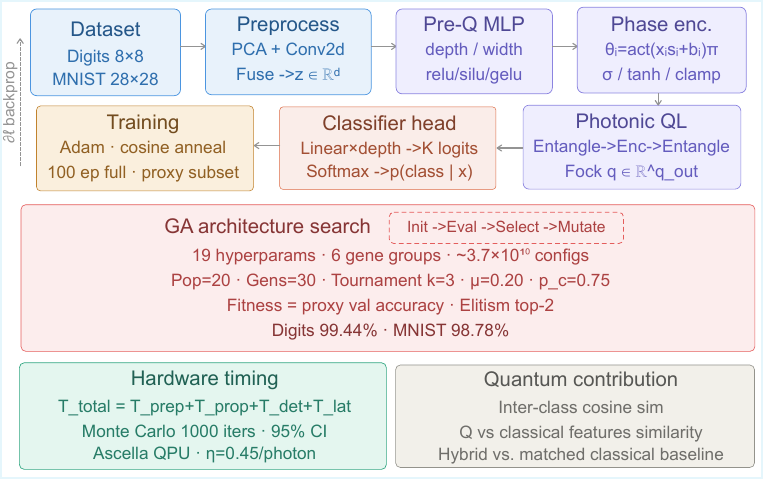}
    \caption{Q-PhotoNAS framework overview. Classical preprocessing (PCA + Conv2d) fuses inputs into $\mathbf{z} \in \mathbb{R}^d$, which passes through a pre-quantum MLP and learnable phase encoder ($\theta_i = \mathrm{act}_\phi(x_i s_i + b_i)\cdot\pi$) before entering the photonic quantum layer (Merlin \texttt{ML.QuantumLayer.simple}~\cite{merlin_layer}). A classical classifier head maps the Fock-basis output $\mathbf{q}$ to class probabilities. A genetic algorithm (pop=20, 30 generations) searches 19 hyperparameters across ${\sim}3.7\times10^{10}$ configurations, optimizing proxy validation accuracy. Hardware execution time is estimated via a first-principles model with Monte Carlo uncertainty quantification on the Quandela Ascella QPU.}
    \label{fig:complete_methodology}
\end{figure}

\subsection{System Overview}
Our system for hybrid photonic quantum-classical learning integrates four phases: data preprocessing, progressive model development, genetic algorithm (GA) based neural architecture search (NAS), and hardware execution time estimation, illustrated in Fig.~\ref{fig:complete_methodology}. Two datasets (Digits~\cite{alpaydin1998digits}, MNIST~\cite{lecun1998gradient}) are first standardized and compressed via Principal Component Analysis (PCA) ~\cite{jolliffe2002pca} to meet quantum hardware constraints. A hybrid model, \texttt{PhotonicModel}, is then constructed around a photonic quantum layer provided by Merlin, Quandela's open-source Python library for building, simulating, and deploying hybrid photonic quantum-classical models, which interfaces directly with the Ascella QPU~\cite{maring2023ascella}, with classical pre and post-processing blocks on either side. To optimize the hyperparameters of this architecture, we use a GA~\cite{holland1975adaptation} that searches across 19 hyperparameters organized into six gene groups, over a search space of approximately $3.7 \times 10^{10}$ configurations. Finally, the execution time of the optimized model is estimated on real photonic hardware using a math-based timing estimator.

\subsection{Data Preprocessing and Dimensionality Reduction}
\label{subsec:Data Preprocessing and Dimensionality Reduction}
Raw input images are processed by two parallel branches before entering the quantum layer. The first branch flattens and standardizes the image using \texttt{StandardScaler}~\cite{scikit-learn}, giving all features zero mean and unit variance, then applies PCA~\cite{jolliffe2002pca} for dimensionality reduction to $d$ components, followed by \texttt{BatchNorm1d} normalization. The second branch is a fixed convolutional frontend that operates directly on the spatial structure of the raw image. It consists of $L=2$ convolutional blocks, each comprising a Conv2d layer (kernel size $k=3$, base channels $C=16$, doubling per block), ReLU activation, BatchNorm2d, and MaxPool2d, followed by adaptive average pooling and a linear projection to $d$ dimensions. The $d$-dimensional outputs of both branches are concatenated into a $2d$-dimensional vector and passed through a fusion layer (Linear($2d \to d$) + SiLU) to produce the unified representation $\mathbf{z} \in \mathbb{R}^d$, which feeds all subsequent stages of the model.

The dimensionality reduction via PCA is motivated by two considerations. First, \texttt{ML.QuantumLayer.simple()}~\cite{merlin_layer} imposes a hard constraint of \texttt{input\_size}~$\leq 20$, so raw pixel features cannot be fed directly to the quantum layer. Second, our experiments show that models trained on PCA-compressed features consistently outperform those trained on raw pixels for the same quantum circuit, as the compression removes noise and focuses learning on the most discriminative components (Fig.~\ref{fig:prog_dev_motivation}). We use $d=8$ components for Digits (from 64 features) and $d=16$ for MNIST (from 784), chosen to balance information retention with hardware constraints. The convolutional frontend complements the PCA branch by preserving local spatial structure that the flattening operation discards, providing the fusion layer with two complementary views of the same input. The frontend is fixed and not part of the GA search space, ensuring that its contribution is consistent across all evaluated architectures.

\begin{figure*}[htbp]
    \centering
    \includegraphics[width=\linewidth]{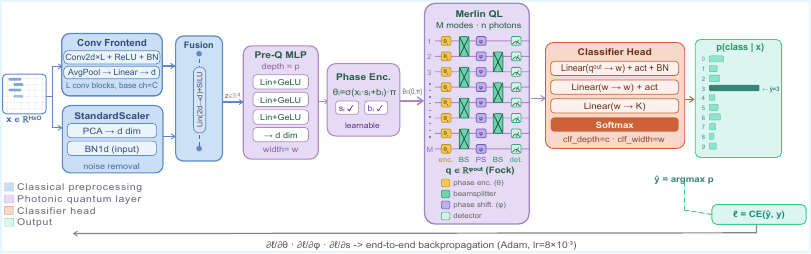}
    \caption{Sample \texttt{PhotonicModel} architecture: full forward pass from raw image $\mathbf{x} \in \mathbb{R}^{H \times W}$ to class prediction. Two parallel branches: (1) Conv Frontend with $L$ blocks (Conv2d+ReLU+BN, base channels $C$), adaptive average pooling, linear projection to $d$; (2) StandardScaler+PCA compresses flattened image to $d$ then BN1d. Concatenate and fuse via Linear($2d\to d$)+SiLU to $\mathbf{z} \in \mathbb{R}^d$. Pre-Q MLP (depth $p$, width $w$, activation \texttt{act}) maps $\mathbf{z}$ higher. Phase Enc.: maps each $x_i$ to phase $\theta_i \in (0,\pi)$ via trainable scale $s_i$, bias $b_i$, activation $\mathrm{act}_\phi \in \{\sigma,\tanh,\mathrm{clamp}\}$. Merlin QL is \texttt{ML.QuantumLayer.simple}~\cite{merlin_layer} photonic circuit with $M$ modes, $n$ photons: encoding applies $\theta$ phases, beam splitter mixes modes, PS ($\phi$) applies trainable phase shift, detector measures probabilities. Output $\mathbf{q} \in \mathbb{R}^{q_\mathrm{out}}$ is Fock-basis vector over $\binom{M+n-1}{n}$ mode-occupation patterns, for $M=(input\_size + 1)$ and $n=\lceil (input\_size + 1) / 2 \rceil$ (the layer returns a ${q_\mathrm{out}}$-dimensional projection/subset of this space, where ${q_\mathrm{out}}$ is a searched hyperparameter)/ Classifier Head (depth $c$, width $w$, activation \texttt{act}, optional BN) maps $\mathbf{q}$ to logits over $K$ classes, then softmax to $p(\mathrm{class}\mid\mathbf{x})$. All parameters $\theta$, $\phi$, $s$ trained end-to-end via Adam ($\mathrm{lr}=\eta$).}
    \label{fig:photonic_arch}
\end{figure*}

\subsection{Learnable Quantum Phase Encoding}
\label{subsec:Learnable Quantum Phase Encoding}
We implement a learnable phase encoding mechanism that replaces fixed, non-adaptive encoding schemes~\cite{senokosov2024encoding,bose2025qdata}. Static encoding maps classical data to quantum phases using a predetermined transformation that cannot be optimized for the task; this limits the expressivity of the quantum layer because the data-to-phase mapping is fixed regardless of what the classifier needs. Since our approach parameterizes the encoding function, it is optimized end-to-end with the rest of the network. The general encoding is:
\begin{equation}
    \theta_i = \mathrm{act}_\phi(x_i \cdot s_i + b_i) \cdot \pi
    \label{eq:encoding}
\end{equation}
where $x_i$ is the $i$-th input feature; $s_i$ is a trainable per-feature scale parameter initialized to \texttt{phase\_scale\_init} $\in \{0.3, 0.5, 0.7, 1.0, 1.5\}$, which controls how broadly the features are spread across the phase range; $b_i$ is an optional trainable per-feature bias initialized to 0.0, included or excluded based on the \texttt{phase\_bias} $\in \{\texttt{True}, \texttt{False}\}$ hyperparameter; and $\mathrm{act}_\phi$ is a searched activation function that determines the shape of the data-to-phase mapping. Both $s_i$ and $b_i$ (when enabled) are learned via backpropagation alongside all other model parameters.

The GA searches over three activation variants for $\mathrm{act}_\phi$, each with distinct properties relevant to the photonic phase space:
\begin{itemize}
    \item \texttt{sigmoid}: maps the scaled input smoothly to $(0,1)$, then scales to $(0,\pi)$. Provides a globally bounded, monotone, and differentiable mapping. Suitable when a soft saturation at both phase extremes is desirable.
    \item \texttt{tanh}: maps to $(-1,1)$, shifted and scaled to $(0,\pi)$, providing a symmetric mapping centred at $\pi/2$. Compresses extreme values more aggressively than sigmoid and may better exploit the full phase range for zero-mean inputs.
    \item \texttt{clamp}: hard-clips the scaled input to $[0, \pi]$ directly, without nonlinear compression. Preserves the linear relationship between input magnitude and phase angle within the valid range, at the cost of zero gradient at saturation.
\end{itemize}
All three variants guarantee output phases in $(0,\pi)$, matching the valid input range of the photonic phase shifters. 

\subsection{Hybrid PhotonicModel Architecture}
\label{subsec:architecture}

\subsubsection{Model Structure}

The \texttt{PhotonicModel} is a four-stage hybrid pipeline, illustrated in Fig.~\ref{fig:photonic_arch}. The first stage is the dual-branch input processing described in Section~\ref{subsec:Data Preprocessing and Dimensionality Reduction}, which produces the fused representation $\mathbf{z} \in \mathbb{R}^d$. The second stage is a classical pre-quantum MLP: a sequence of \texttt{pre\_depth} hidden layers, each consisting of a Linear transformation to width \texttt{pre\_width}, an activation function \texttt{pre\_activation}, and optional dropout \texttt{pre\_dropout}, followed by an optional \texttt{BatchNorm1d}~\cite{ioffe2015batch} (\texttt{pre\_bn} and a final Linear projection back to $d$. When \texttt{pre\_depth}$=0$, this stage reduces to the single final projection, bypassing the hidden layers entirely. The pre-quantum MLP extracts higher-level representations from $\mathbf{z}$ before quantum encoding. The third stage is the photonic quantum layer, described below. The fourth stage is the classical classifier head.

The photonic layer is instantiated as \texttt{ML.QuantumLayer.simple(input\_size=$d$)}~\cite{merlin_layer}, which builds a linear-optical circuit over $d$ modes consisting of a fully trainable entangling layer, a phase-encoding layer that injects the $\theta_1,\ldots,\theta_d$ values produced by the phase encoder, and a second trainable entangling layer. Each entangling layer uses beamsplitter and phase-shifter components. The GA searches over \texttt{q\_output\_size}, which controls the dimensionality $q_\mathrm{out}$ of the quantum output vector; when set to \texttt{None}, the output dimension equals the input size $d$. The layer output $\mathbf{q} \in \mathbb{R}^{q_\mathrm{out}}$ is a photon-number probability vector over the Fock basis: the set of all distinct ways to distribute $n$ photons across $d$ modes, where each basis state is labeled by the occupation number of each mode (such as state $(2,1,0,\ldots)$ means 2 photons in mode 0, 1 in mode 1, etc.). The layer is fully differentiable and trained end-to-end through Merlin's simulation backend.

The classifier head maps $\mathbf{q}$ to class logits via \texttt{clf\_depth} Linear layers of width \texttt{clf\_width}, with activation \texttt{clf\_activation}, \texttt{gelu}\}, optional \texttt{BatchNorm1d} (\texttt{clf\_bn}, and dropout \texttt{clf\_dropout} applied to all but the final layer, which projects to $K$ class logits followed by softmax.

\subsubsection{Search Space and Genome Design}

\begin{figure}[t!]
    \centering
\includegraphics[width=\columnwidth]{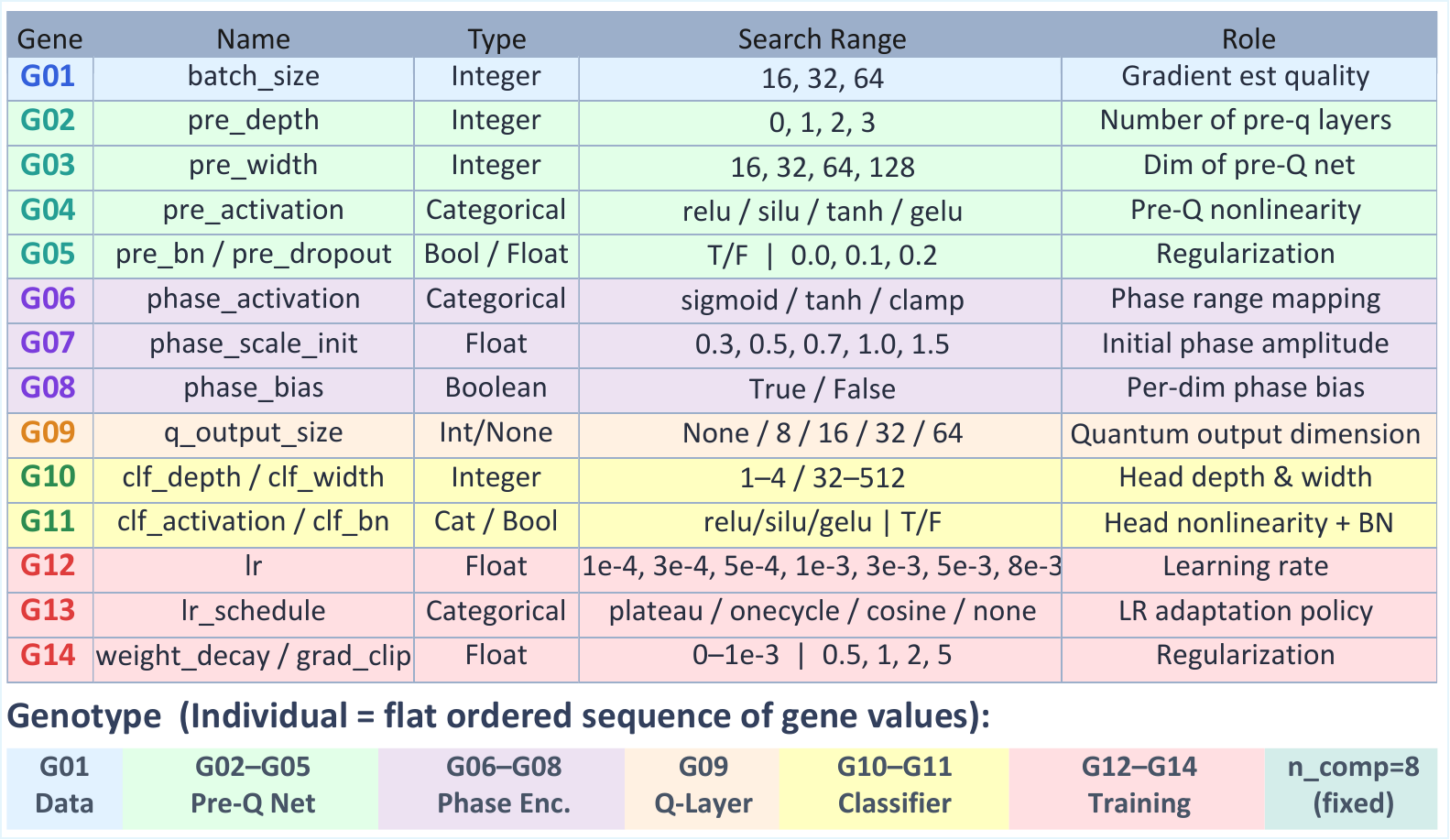}
    \caption{Genome structure. Each individual encodes 19 genes across six functional groups. Genes within a group are inherited together during crossover.}
    \label{fig:genome_structure}
\end{figure}

The search space covers 19 hyperparameters spanning approximately $3.7 \times 10^{10}$ configurations, organized into six functional gene groups shown in Fig.~\ref{fig:genome_structure}.Grouping is essential because hyperparameters within a group are co-adapted; for example, \texttt{pre\_depth} and \texttt{pre\_width} should come from the same parent, since inheriting a large depth from one parent and a small width from another produces an inconsistent architecture~\cite{white2023nas}.

\begin{figure*}[htbp]
    \centering
    \includegraphics[width=\linewidth]{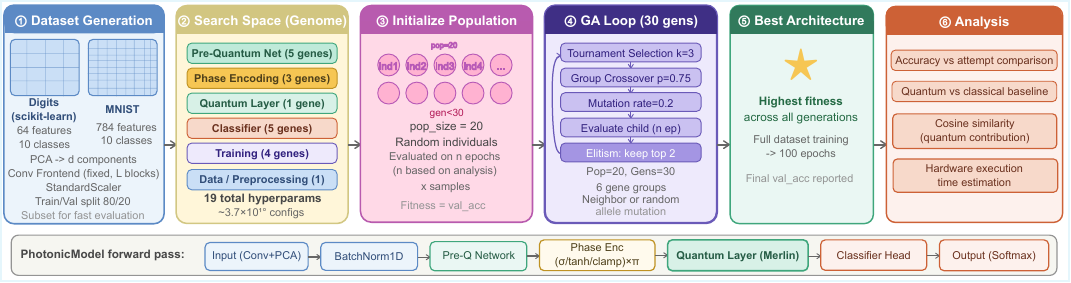}
    \caption{GA-based architecture search methodology. From dataset and search space through population evolution to the best architecture and analysis.}
    \label{fig:ga_loop}
\end{figure*}

\subsection{GA-Based Neural Architecture Search}
To efficiently explore the hyperparameter space of \texttt{PhotonicModel}, we implement a GA~\cite{holland1975adaptation} designed for hybrid quantum-classical architectures (Algorithm~\ref{algo:architecture_search}).  The search loop, shown in Fig.~\ref{fig:ga_loop}, evolves a population of 20 candidate architectures over 30 generations, evaluating each by its validation accuracy after a short training budget.

\begin{algorithm}[t!]
\caption{GA-Based Architecture Search}
\small
\label{algo:architecture_search}
\KwData{Population size $N=20$, generations $G=30$}
\KwResult{Best architecture}
Initialize population of $N$ random individuals\;
\For{$g \gets 1$ \KwTo $G$}{
    Evaluate fitness of each individual\;
    Select parents via tournament selection ($k=3$)\;
    Generate offspring via group crossover ($p_c=0.75$) and per-gene mutation ($\mu=0.20$)\;
    Apply elitism: carry top 2 individuals unchanged\;
    Form next generation\;
}
\Return individual with highest fitness across all generations\;
\end{algorithm}
 
\subsubsection{Evolutionary Operators}

\begin{figure}[t!]
    \centering
    \includegraphics[width=\columnwidth]{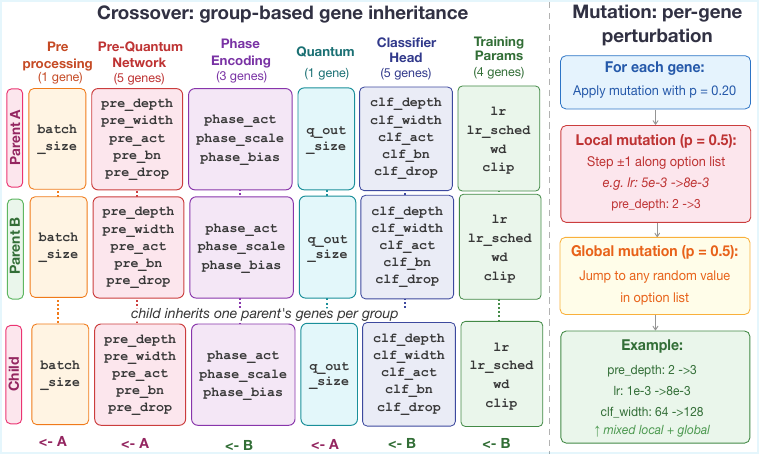}
    \caption{Group-based crossover example. Each group is chosen independently, giving 64 possible combinations per crossover.}
    \label{fig:group_crossover}
\end{figure}

\begin{figure}
    \centering
    \includegraphics[width=0.7\columnwidth]{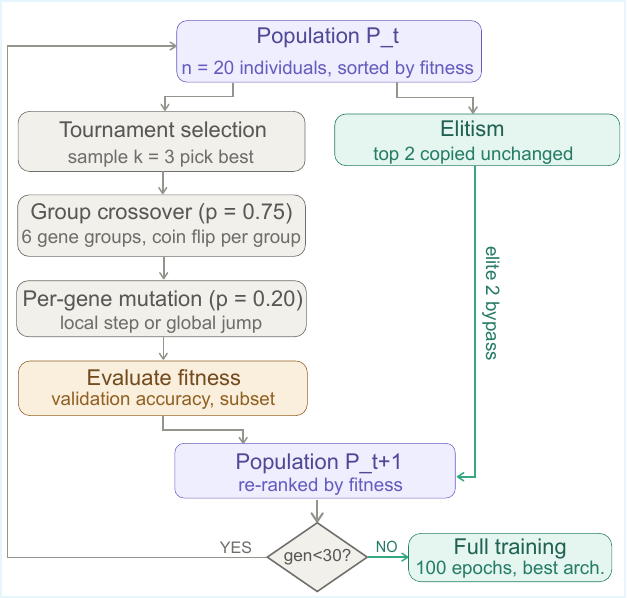}
    \vspace{5pt}
    \caption{One generation of the GA: tournament selection and elitism produce parents; crossover and mutation produce offspring; each offspring is evaluated by short training; the population is re-ranked for the next generation.}
    \label{fig:ga_iteration}
\end{figure}

The GA uses four mechanisms, illustrated in Figs.~\ref{fig:group_crossover} and~\ref{fig:ga_iteration}, and described in Algorithm~\ref{algo:group_crossover}.

\textbf{Tournament selection} ($k=3$)~\cite{miller1995tournament}: three individuals are drawn uniformly at random from the population and the one with the highest fitness is selected as a parent. This balances exploitation of high-fitness individuals with exploration by maintaining population diversity. Larger $k$ increases selection pressure; $k=3$ is a common moderate-pressure setting.

\textbf{Group-based crossover} ($p_c=0.75$): with probability $p_c$ two parents are recombined; otherwise the child is a direct copy of the selected parent. When recombination occurs, each of the six gene groups is inherited as a unit from one parent, chosen independently with probability 0.5. This yields $2^6 = 64$ distinct offspring combinations per crossover event. Unlike single-point crossover, which splits the genome at a fixed position and can separate co-adapted parameters, group crossover preserves architectural coherence at component boundaries. The operator is visualized in Fig.~\ref{fig:group_crossover}.

\textbf{Per-gene mutation} ($\mu=0.20$ per gene): each gene is mutated independently with probability $\mu$. When a gene mutates, it has a 50\% chance of a local step (moving one position along its ordered option list, clamped to the list boundaries) and a 50\% chance of a global jump to a uniformly random value from its option list. Local steps incrementally refine promising configurations; global jumps help escape local optima and maintain diversity~\cite{vincent2023automl}.

\textbf{Elitism}: the top 2 individuals by fitness are copied unchanged in the next generation, guaranteeing a monotonic improvement in the best-observed fitness across generations~\cite{holland1975adaptation}.

\begin{algorithm}[t!]
\caption{Group-Based Crossover}
\label{algo:group_crossover}
\small
\KwData{Parents $a, b$, crossover probability $p_c=0.75$}
\KwResult{Child individual}
\eIf{random $< p_c$}{
    Initialize empty child\;
    \For{each gene group}{
        \eIf{random $< 0.5$}{
            inherit group from $a$\;
        }{
            inherit group from $b$\;
        }
    }
}{
    child $\gets$ copy of $a$\;
}
\Return child\;
\end{algorithm}
\subsubsection{Fitness Evaluation}
Each candidate is evaluated on a short proxy budget (5 epochs/1000 samples for Digits; 3 epochs/5000 samples for MNIST); the best architecture found after 30 generations is then fully retrained for 100 epochs on the complete dataset using the schedule determined by the GA search.

\subsection{Proxy Training Budget Selection via Correlation Analysis}
\label{subsec:epoch_selection}

To determine the minimum reliable proxy budget, we correlate checkpoint accuracy at each intermediate epoch against final accuracy (epoch~20) across 24 randomly sampled architectures~\cite{maleki2026qnasneuralarchitecturesearch}, computing both Pearson~($r$) and Spearman~($\rho$) coefficients on the same evaluation subsets used during GA search. Spearman~$\rho$ is the primary metric since the GA requires only correct ranking, not accurate absolute fitness prediction. Full results are in Section~\ref{subsec:epoch_results}.

\subsection{Impact on Quantum Component}
To quantify the photonic layer's contribution, we compute two complementary metrics on the fully trained model. First, \textit{inter-class cosine similarity} of the quantum output vectors: for each pair of digit classes, we compute the cosine similarity between their mean Fock-basis output vectors; a low mean value indicates that the photonic layer maps different classes to well-separated, near-orthogonal regions of its output space, which directly aids downstream classification. Second, \textit{quantum-classical feature orthogonality}: the per-sample cosine similarity between the quantum layer's output and the pre-quantum classical features entering it; a value near zero indicates the photonic layer introduces genuinely new information rather than replicating what the classical pathway already encodes, while a negative value indicates active anti-alignment, a stronger form of complementarity. Together, these two metrics distinguish between a photonic layer that separates classes well internally and one that adds information the classical stream cannot provide; both are necessary conditions for a meaningful quantum contribution. We also compare the hybrid model against a matched classical-only baseline with the quantum layer removed and parameter budget equalized; full results are reported in Section~\ref{sec:quantum_contribution_results}.

\subsection{Hardware Execution Time Estimation}
\begin{figure}[t!]
    \centering
    \includegraphics[width=\columnwidth]{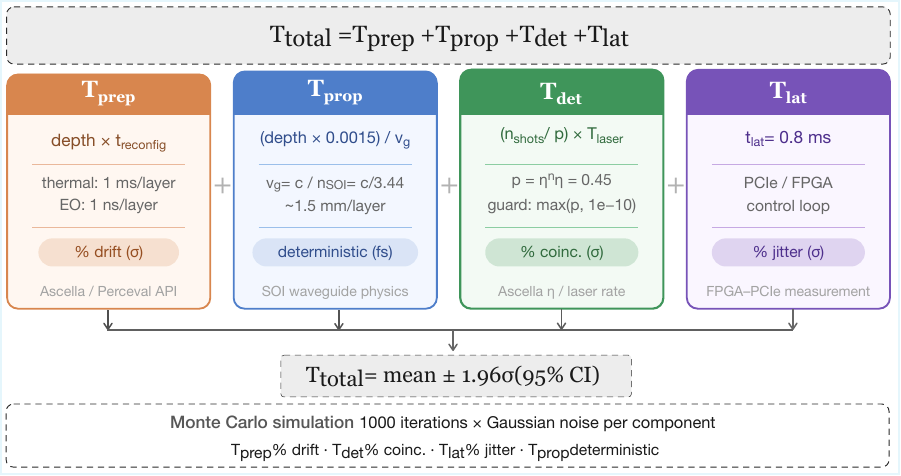}
    \vspace{5pt}
    \caption{Hardware execution time estimator: first-principles mathematical model with Monte Carlo confidence interval. $T_\text{prep}$ is the thermal phase-shifter reconfiguration time, subject to 2\% thermal drift noise. $T_\text{prop}$ is the photon propagation time through silicon-on-insulator (SOI) waveguides (picosecond scale, deterministic). $T_\text{det}$ is the detection and sampling time accounting for photon loss, with a $p_\text{success}$ underflow guard and a dedicated 10\% photon-coincidence uncertainty in the confidence interval. $T_\text{lat}$ is the fixed PCIe/FPGA electronic control overhead, subject to 5\% jitter noise. The 95\% CI is computed from 1000 Monte Carlo iterations.}
    \label{fig:hardware_estimator}
\end{figure}

To assess real-world viability, we estimate the execution time of the hybrid model on the Quandela Ascella QPU~\cite{maring2023ascella}using a first-principles mathematical model, illustrated in Fig.~\ref{fig:hardware_estimator}:
\begin{equation}
    T_{\text{total}} = T_{\text{prep}} + T_{\text{prop}} + T_{\text{det}} + T_{\text{lat}}
    \label{eq:hw}
\end{equation}
where $T_{\text{prep}}$ is the phase-shifter reconfiguration time; the time required to thermally heat the phase shifters to apply a new set of phase values, computed as circuit depth $\times$ a per-layer reconfiguration constant (sourced from Ascella QPU specifications~\cite{maring2023ascella,parra2024tops}); $T_{\text{prop}}$ is the light propagation time through the silicon waveguides, computed from the group velocity $v_g = c/n_\text{SOI}$, where $c$ is the speed of light in vacuum and $n_\text{SOI}=3.44$ is the refractive index of silicon-on-insulator waveguides~\cite{dulkeith2006groupindex}; $T_{\text{det}}$ is the detection and sampling time, scaled by the expected number of shots needed to compensate for photon loss; the probability that all $n_\text{photons}$ photons survive the circuit is $p_\text{success} = \eta^{n_\text{photons}}$, where $\eta=0.45$ is the per photon transmission efficiency of the Ascella device~\cite{maring2023ascella}, with a numerical guard $p_\text{success} \geq 10^{-10}$ to prevent divergence at high photon counts; and $T_{\text{lat}} = 0.8\,\text{ms}$ is fixed PCIe/FPGA control loop overhead~\cite{maring2023ascella}. Monte Carlo simulation (1000 iterations) propagates independent Gaussian noise over three components: 2\% thermal phase drift on $T_{\text{prep}}$, a dedicated 10\% photon-coincidence probability uncertainty on $T_{\text{det}}$~\cite{parra2024tops}, and 5\% electronic jitter on $T_{\text{lat}}$; $T_{\text{prop}}$ is treated as deterministic at femtosecond precision. This yields a 95\% confidence interval via $\text{mean} \pm 1.96\sigma$.
\section{Results and Discussion}
\label{sec:results}
 
\subsection{Experimental Setup}
All experiments run on CPU. The GA is configured with population size 20, 30 generations, crossover rate 0.75, mutation rate 0.20, elitism top-2, and tournament size $k=3$~\cite{miller1995tournament}. Each individual is evaluated by training for 5 epochs on a fixed 1000-sample (Digits~\cite{alpaydin1998digits}) or 3 epochs on a fixed 5000-sample (MNIST~\cite{lecun1998gradient}) subset, over the search space outlined in Fig.~\ref{fig:genome_structure}. The best architecture found is then fully trained for 100 epochs on the complete dataset using the Adam optimizer~\cite{kingma2015adam}; the Digits architecture uses cosine annealing~\cite{loshchilov2017sgdr} with \texttt{ReduceLROnPlateau}, while the MNIST architecture uses a one-cycle learning rate schedule, as determined by the GA search. Datasets and splits follow scikit-learn~\cite{scikit-learn} defaults for Digits and the standard train/test split for MNIST. PCA~\cite{jolliffe2002pca} is set to 8 components for Digits and 16 for MNIST. All multi-seed comparisons (classical vs.\ hybrid) are reported as the mean $\pm$ standard deviation over three independent random seeds.

\subsection{Proxy Training Budget Validation}
\label{subsec:epoch_results}

\begin{figure}[t!]
    \centering
    \includegraphics[width=\columnwidth, page=3]{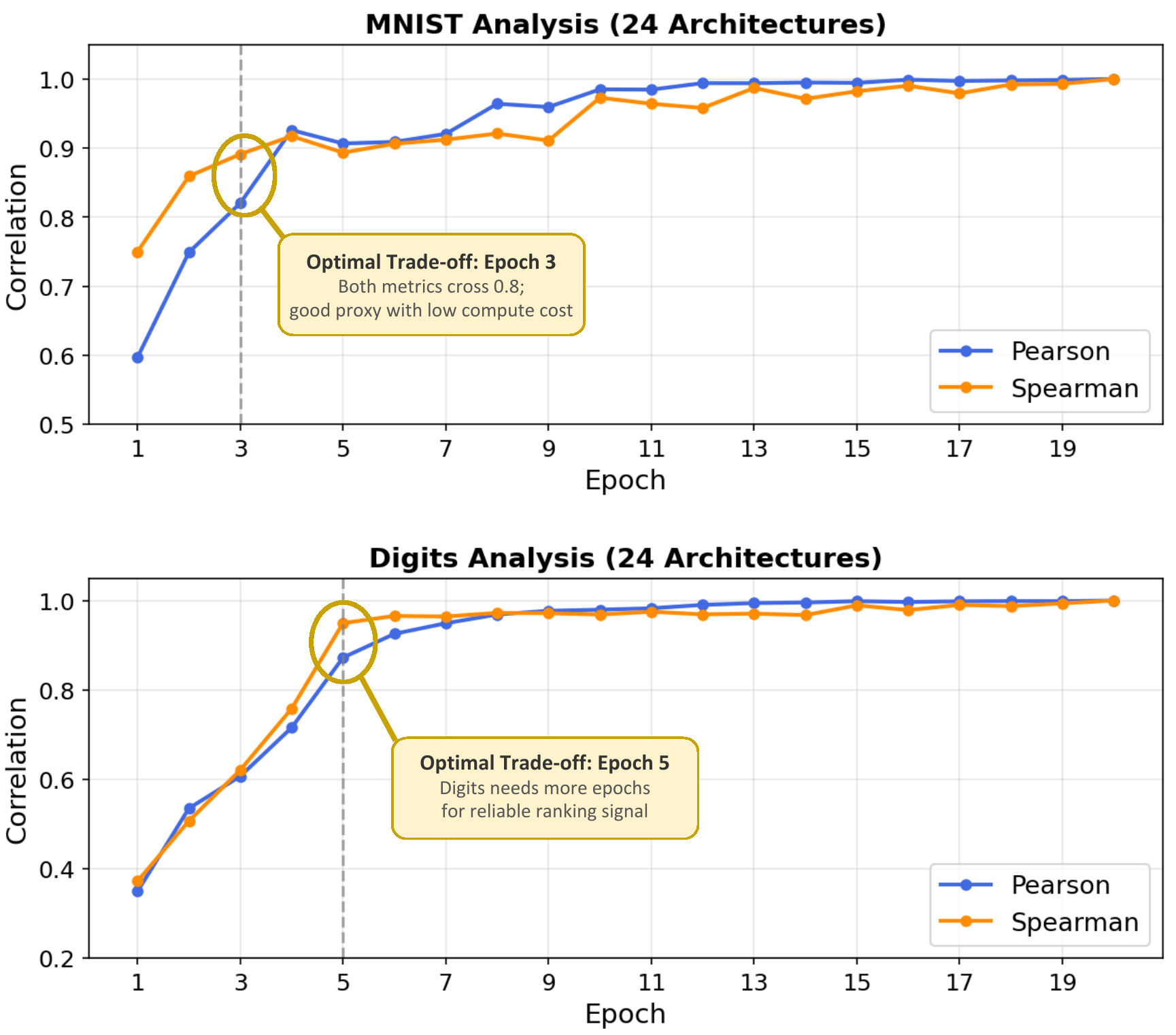}
    \caption{Epoch-selection correlation analysis across 24 randomly sampled architectures. Pearson~$r$ (blue) and Spearman~$\rho$ (orange) correlations between checkpoint accuracy and final (epoch~20) accuracy are shown for MNIST (top, 5000-sample subset) and Digits (bottom, 1000-sample subset). The dashed vertical lines mark the first epoch where both coefficients exceed 0.8: epoch~3 for MNIST and epoch~5 for Digits. Both correlations plateau above 0.95 shortly after, validating the use of short proxy budgets during GA search.}
    \label{fig:epoch_correlation}
\end{figure}

Fig.~\ref{fig:epoch_correlation} reports the correlation analysis described in Section~\ref{subsec:epoch_selection}. The two datasets exhibit notably different convergence profiles. On MNIST, rankings stabilize rapidly: Spearman~$\rho$ already reaches 0.859 at epoch~2 and both metrics cross 0.8 at epoch~3, reflecting the larger and more diverse evaluation subset (5000 samples) which provides low-variance fitness estimates even under very short training. On Digits, convergence is more gradual, with both metrics first exceeding 0.8 at epoch~5; the smaller subset (1000 samples from a 1797-sample dataset) introduces higher fitness variance in early epochs, requiring slightly more training for rankings to stabilize. Notably, the Spearman coefficient on Digits jumps from 0.758 at epoch~4 to 0.950 at epoch~5, a sharp transition suggesting that epoch~5 is the point at which most architectures have escaped random initialization and their relative ordering reflects genuine capacity differences rather than initialization noise. Beyond epoch~5 for Digits and epoch~3 for MNIST, both metrics plateau above 0.95 and climb steadily toward 1.0, confirming that additional epochs refine but do not materially change architecture rankings. These results highlight an important asymmetry: smaller datasets require a larger epoch budget relative to dataset size for stable rankings, while larger datasets stabilize quickly.

\subsection{Algorithmic Convergence Performance}
 
\begin{figure*}[t!]
    \centering
    \includegraphics[width=\linewidth]{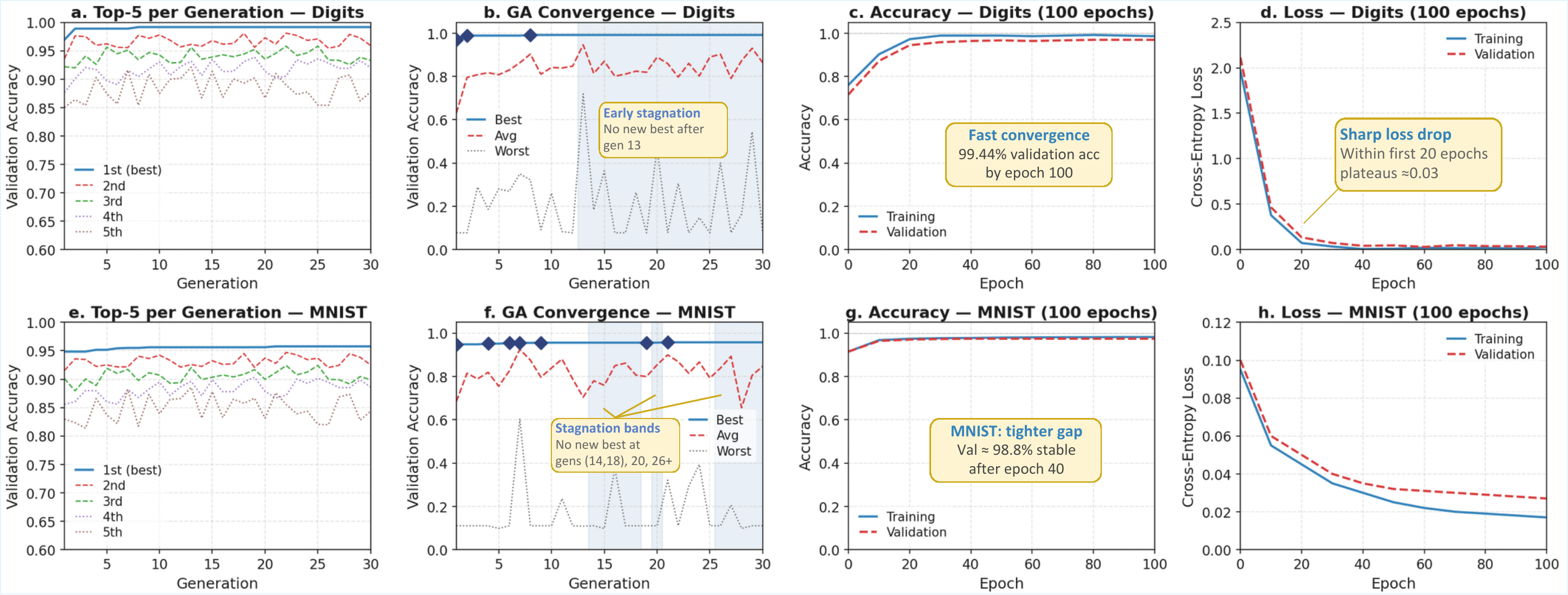}
    \caption{Summary of training and GA search results. \textbf{(a)} Validation accuracy over 100 epochs for the best Digits architecture; final val\_acc = 99.44\%. \textbf{(b)} Cross-entropy loss over 100 epochs for Digits; training and validation losses converge closely, indicating no over-fitting. \textbf{(c)} Top-5 individual fitness per generation during GA search on Digits; the best individual (solid line) reaches 0.9917 at generation~8. \textbf{(d)} GA convergence on Digits: best, average, and worst fitness over 30 generations; diamond markers indicate new best events; shaded regions highlight stagnation periods. \textbf{(e)} Validation accuracy over 100 epochs for the best MNIST architecture; steady improvement from 95.46\% to 98.78\%. \textbf{(f)} Cross-entropy loss over 100 epochs for MNIST; loss falls sharply in the first 20 epochs and plateaus with training and validation tracks closely aligned. \textbf{(g)} Top-5 individual fitness per generation during GA search on MNIST; convergence is slower and noisier than Digits, reflecting a more complex fitness landscape with extended stagnation from generations \textbf{(h)} GA convergence on MNIST: best fitness reaches 0.9577 at generation~21 and plateaus through generation~30.}
    \label{fig:results_combined}
\end{figure*}

\subsubsection{Digits Dataset}

\begin{table}[htbp]
\centering
\caption{Best architectures found by GA for each dataset}
\label{tab:best_arch}
\begin{tabular}{lcc}
\toprule
Hyperparameter & Digits & MNIST \\
\midrule
\texttt{batch\_size}        & 16      & 16       \\
\texttt{pre\_depth}         & 0       & 0        \\
\texttt{pre\_width}         & 16      & 128      \\
\texttt{pre\_activation}    & silu    & tanh     \\
\texttt{pre\_bn}            & False   & True     \\
\texttt{pre\_dropout}       & 0.2     & 0.2      \\
\texttt{phase\_activation}  & tanh    & clamp    \\
\texttt{phase\_scale\_init} & 1.5     & 1.5      \\
\texttt{phase\_bias}        & False   & True     \\
\texttt{q\_output\_size}    & 16      & 64       \\
\texttt{clf\_depth}         & 3       & 2        \\
\texttt{clf\_width}         & 64      & 64       \\
\texttt{clf\_activation}    & silu    & silu     \\
\texttt{clf\_bn}            & True    & True     \\
\texttt{clf\_dropout}       & 0.0     & 0.3      \\
\texttt{lr}                 & 8e-3    & 5e-4     \\
\texttt{lr\_schedule}       & cosine  & onecycle \\
\texttt{weight\_decay}      & 1e-4    & 0.0      \\
\texttt{grad\_clip}         & 1.0     & 1.0      \\
\midrule
GA best fitness               & 0.9917  & 0.9577   \\
Final val\_acc (full dataset) & 0.9944  & 0.9878   \\
\bottomrule
\end{tabular}
\end{table}

Figs.~\ref{fig:results_combined}a and~\ref{fig:results_combined}b show the GA convergence over 30 generations on Digits. The population improves steadily in the early generations, reaching an initial best fitness of 0.9694 at generation~1 and accumulating three new-best improvements to reach a final best of 0.9917 at generation~8. Stagnation sets in from generation~13 onward with no further improvements across the remaining 22 generations, indicating the search has closely approached the effective optimum of the landscape within the allotted budget. The average fitness tracks behind the best throughout, rising from 0.6317 at generation~1 to 0.8637 at generation~30, reflecting a healthy mix of exploration and exploitation. The final architecture outlined in Table~\ref{tab:best_arch} achieves 99.17\% fitness on the 1000-sample evaluation subset using a 5-epoch proxy budget. Full training yields a val\_acc of 99.44\% (Fig.~\ref{fig:results_combined}c-d), with training and validation losses tightly aligned throughout, confirming no overfitting.

\subsubsection{MNIST Dataset}

Figs.~\ref{fig:results_combined}e and~\ref{fig:results_combined}f show GA convergence on MNIST over 30 generations. The best fitness improves from 0.9485 at generation~1 to 0.9577 at generation~21, accumulating seven new-best improvements across the run. Stagnation dominates the final nine generations (gens~22-30 show no further improvement), suggesting the search has converged within the available evaluation budget. The average fitness on MNIST is noisier than Digits, oscillating between 0.66 and 0.90 across generations, reflecting a more complex and less smooth fitness landscape consistent with MNIST~\cite{lecun1998gradient} being a harder task with greater intraclass variability and more architectural sensitivity. Notably, an extended stagnation window appears at generations~14-18, before a late breakthrough at generation~19 restarts improvement through generation~21, after which the search plateaus through generation~30. This pattern is consistent with the shorter 3-epoch proxy budget offering higher fitness variance and a less smooth selection signal than the 5-epoch budget used for Digits. The best architecture (Table~\ref{tab:best_arch}) achieves 95.77\% on the 5000-sample evaluation subset under the 3-epoch proxy. Full training on the complete 60k MNIST dataset for 100 epochs yields a best validation accuracy of 98.78\% (Figs.~\ref{fig:results_combined}g and~\ref{fig:results_combined}h), Full training on the complete 60k dataset yields 98.78\%, with the 3.0\,pp proxy gap (vs.\ 0.27\,pp for Digits) analyzed in Section~\ref{sec:discussion}.

\subsection{Hardware Estimation}

\begin{table}[t!]
\centering
\vspace{5pt}
\caption{Per-image hardware execution time on the Quandela Ascella thermal QPU. Values are mean $\pm$ std over 1000 Monte Carlo iterations (2\% thermal drift, 10\% coincidence uncertainty, 5\% electronic jitter).}
\label{tab:hw_latency}
\begin{adjustbox}{max width=\linewidth}
\begin{tabular}{lcc}
\toprule
Component & Digits (9 modes, depth 65) & MNIST (17 modes, depth 129) \\
\midrule
$T_{\text{prep}}$ (phase-shifter reconfig.) & 65.0000\,ms & 129.0000\,ms \\
$T_{\text{prop}}$ (waveguide propagation)   & $\approx 0$\,ms & $\approx 0$\,ms \\
$T_{\text{det}}$  (detection \& sampling)   & 0.6774\,ms & 16.5195\,ms \\
$T_{\text{lat}}$  (PCIe/FPGA overhead)      & \multicolumn{2}{c}{0.8000\,ms} \\
\midrule
Quantum subtotal   & $66.550 \pm 2.468$\,ms & $146.453 \pm 5.762$\,ms \\
Classical subtotal & $0.535 \pm 0.650$\,ms  & $2.159 \pm 0.423$\,ms \\
\midrule
\textbf{Total per-image latency} & $\mathbf{67.09 \pm 2.55}$\,\textbf{ms} & $\mathbf{148.61 \pm 5.78}$\,\textbf{ms} \\
\bottomrule
\end{tabular}
\end{adjustbox}
\end{table}

Table~\ref{tab:hw_latency} summarizes the per-image latency on the Quandela Ascella QPU; Analysis of the dominant cost components is in Section~\ref{sec:discussion}.

\subsection{Quantum Contribution Analysis}
\label{sec:quantum_contribution_results}

\begin{figure}[t!]
    \centering
    \includegraphics[width=\columnwidth]{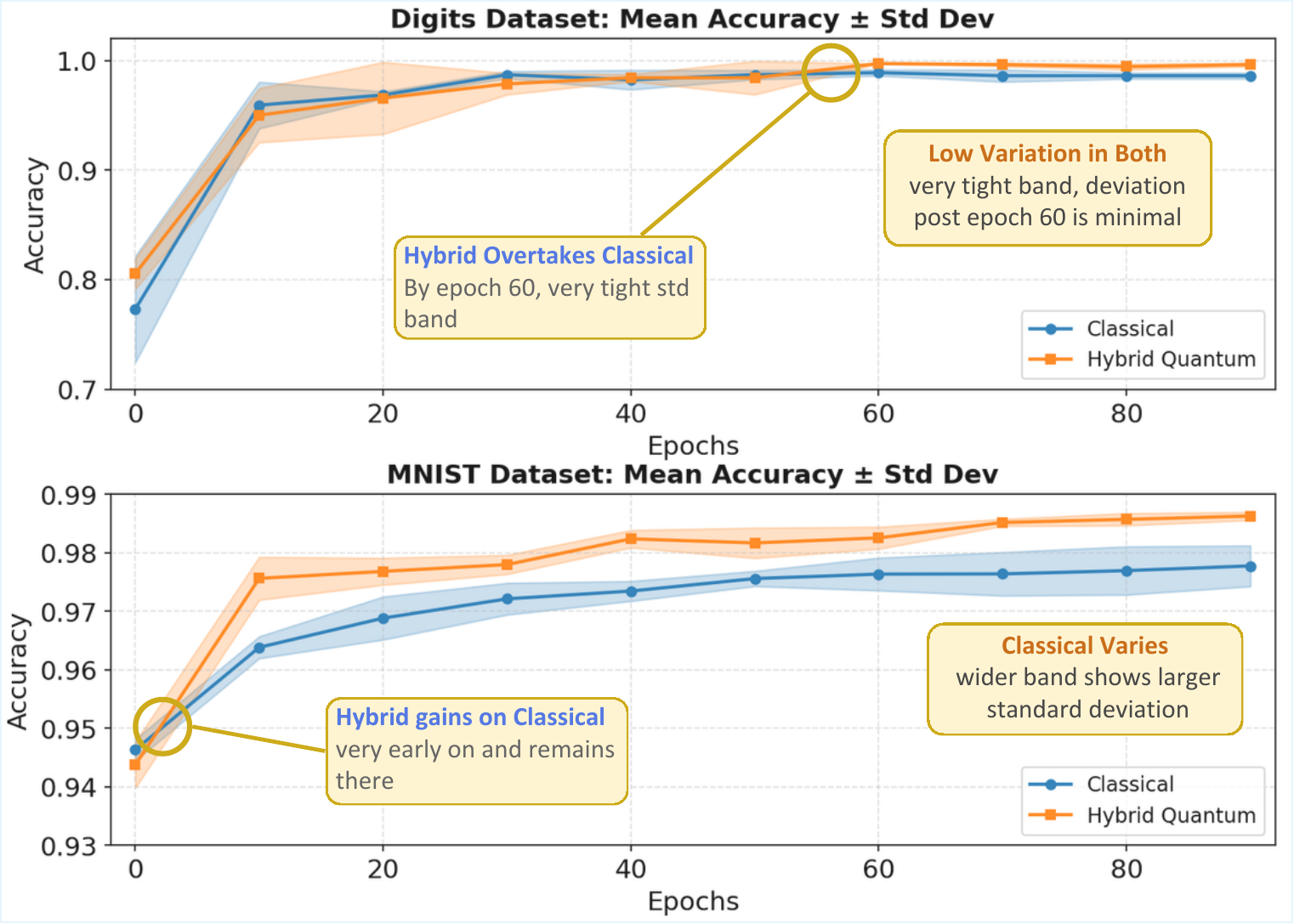}
    \caption{Training accuracy comparison between classical-only and hybrid quantum-classical models on Digits (up) and MNIST (down) datasets. Solid lines represent the mean accuracy across three independent seeds; shaded regions indicate one standard deviation. On Digits the hybrid model maintains a consistent accuracy advantage throughout training. On MNIST, the hybrid model's advantage is similarly consistent, reaching a mean best accuracy of 98.75\% versus 97.92\% for the classical baseline.}
    \label{fig:quantum_analysis}
\end{figure}

To isolate the contribution of the photonic layer we compare the full hybrid model against a matched classical-only baseline (same architecture with the quantum layer removed and parameter budget equalized; 11,658 parameters for Digits and 11,706 for MNIST) over three seeds, shown in Fig.~\ref{fig:quantum_analysis}. On Digits, the hybrid model achieves a mean best val\_acc of 99.81\% across seeds, compared to the classical baseline at 99.16\%, an advantage of approximately 0.65\,pp. On MNIST, the hybrid model achieves a mean best val\_acc of 98.75\%, compared to 97.92\% for the classical baseline, an advantage of approximately 0.83\,pp. Notably, the hybrid model also exhibits substantially lower variance than the classical baseline on both datasets, suggesting that the photonic layer acts as a stabilising inductive bias in addition to improving mean accuracy. 

For the GA-found \texttt{PhotonicModel} trained on Digits, the mean inter-class cosine similarity of the quantum layer output vectors across the 10 digit classes is 0.135 ($\sigma = 0.082$). The near-zero mean confirms that the quantum layer maps different digit classes to largely orthogonal regions of the Fock-basis output space. The per-sample cosine similarity between the quantum output and the pre-quantum classical features has a mean of 0.167 ($\sigma = 0.072$), indicating that the photonic layer introduces information not already present in the classical pathway. Note that Fock-basis probability vectors are non-negative and sum to one, which geometrically constrains achievable cosine similarities to the positive range and compresses the distribution toward zero; the observed mean of 0.135 should therefore be interpreted relative to this structural constraint rather than against a $[-1, +1]$ baseline. For the MNIST architecture, the mean inter-class cosine similarity of the quantum outputs is 0.100 ($\sigma = 0.086$), indicating broadly consistent separation across all 45 class pairs. The per-sample cosine similarity between the quantum output and the classical pre-quantum features has a mean of $-$0.154 ($\sigma = 0.113$), confirming that the quantum layer not only extracts information orthogonal to the classical pathway but actively contrasts it; unlike Digits, the photonic output on MNIST is geometrically anti-aligned with the pre-quantum features, a stronger form of complementarity that is not subject to the non-negativity constraint on the inter-class metric and suggests the phase encoding and interference pattern learned for MNIST genuinely inverts aspects of the classical representation.
 
\subsection{Discussion}
\label{sec:discussion}

\subsubsection{Dataset-Specific Architecture Preferences}

The two best architectures in Table~\ref{tab:best_arch} reveal a clear pattern of shared structure and task-specific adaptation. Both converge on \texttt{pre\_depth=0}, \texttt{clf\_width=64}, \texttt{clf\_activation=silu}, \texttt{clf\_bn=True}, \texttt{batch\_size=16}, \texttt{phase\_scale\_init=1.5}, and \texttt{grad\_clip=1.0}, reflecting general principles for stable gradient flow in this hybrid photonic setting. The differences are equally informative: MNIST selects a substantially wider pre-quantum network (\texttt{pre\_width=128} vs.\ 16), a larger quantum output dimension (\texttt{q\_output\_size=64} vs.\ 16), a \texttt{clamp} phase activation rather than \texttt{tanh}, batch normalisation before the quantum layer (\texttt{pre\_bn=True}), and stronger classifier regularisation (\texttt{clf\_dropout=0.3} vs.\ 0.0). These differences are consistent with MNIST's greater intra-class variability and higher-dimensional PCA input (16 components vs.\ 8), both of which benefit from richer feature transformation before and after the photonic layer. The learning rate regimes also diverge: Digits is best served by an aggressive cosine-annealed schedule starting at $lr=8\times10^{-3}$, whereas MNIST favours a conservative one-cycle policy at $lr=5\times10^{-4}$ with $L_2$ regularisation disabled, suggesting that the larger dataset already provides sufficient implicit regularisation through data diversity.

\subsubsection{GA Proxy Budget and Full-Dataset Accuracy}

The gap between GA fitness and final full-dataset accuracy differs between datasets in a predictable and informative way. For Digits, the GA reports 99.17\% under a 5-epoch proxy on a 1000-sample subset, while full training on the complete dataset for 100 epochs yields a best validation accuracy of 99.44\%, a modest 0.27\,pp underestimate that closes rapidly under cosine annealing. For MNIST, the gap is considerably larger: 95.77\% (GA proxy on 5000 samples, 3 epochs) vs.\ 98.78\% (full 60k dataset, 100 epochs), approximately 3.0\,pp. This larger gap reflects both the shorter relative proxy budget (3 epochs vs.\ 5) and the substantial additional gradient signal available from the full 60k training set that is absent during proxy evaluation. Even though the proxy underestimates absolute fitness, architecture rankings are preserved before accuracy values converge, so the selection pressure remains unbiased, the desirable regime for proxy-based NAS~\cite{white2023nas}. The tighter proxy gap on Digits (0.27\,pp vs.\ 3.0\,pp) is consistent with the smaller dataset size and lower task complexity: a 1000-sample proxy already captures most of the accuracy signal available in the full 1797-sample Digits corpus, whereas a 5000-sample proxy covers only 8.3\% of the 60k MNIST training set, leaving substantial room for accuracy gains during full training.

\subsubsection{Hardware Viability and Dominant Cost Analysis}

The hardware estimation results in Table~\ref{tab:hw_latency} reveal that the photonic quantum component dominates total latency in both architectures: the quantum subtotal accounts for 99.2\% of total latency on Digits ($66.55\,\text{ms}$ of $67.09\,\text{ms}$) and 98.5\% on MNIST ($146.45\,\text{ms}$ of $148.61\,\text{ms}$). The classical wrapper is negligible in both cases ($0.535\,\text{ms}$ for Digits, $2.159\,\text{ms}$ for MNIST), indicating that optimizing the classical components yields no meaningful latency reduction for current configurations. Within the quantum budget, thermal phase-shifter reconfiguration ($T_{\text{prep}}$) accounts for 97.7\% and 88.1\% of the quantum cost for Digits and MNIST respectively, scaling directly with circuit depth (65 for Digits, 129 for MNIST). This makes circuit depth the primary lever for latency optimization. A transition to electro-optic phase shifters, which operate at nanosecond rather than millisecond timescales~\cite{parra2024tops}, would reduce $T_{\text{prep}}$ by several orders of magnitude and shift the bottleneck to detection. On current thermal hardware, single-image inference in the 67-149\,ms range is feasible and consistent with near-term deployment scenarios where throughput requirements are moderate. The detection component ($T_{\text{det}}$) is not negligible for MNIST (16.52\,ms, 11.3\% of quantum budget), driven by the larger photon number (9 photons) and the exponential scaling of shot requirements with photon loss ($p_{\text{success}} = \eta^{n_{\text{photons}}}$ with $\eta = 0.45$). This highlights the importance of photon transmission efficiency as a hardware target for deeper photonic circuits.

\subsubsection{Limitations and Future Research Directions}

While our results demonstrate the effectiveness of GA-based NAS for hybrid photonic architectures, several directions remain open and represent natural extensions of this work. The current framework targets simulation based evaluation via Merlin's backend; deploying the optimized architectures on physical Ascella hardware would provide empirical validation of the timing model and expose noise characteristics that simulation may underestimate. Furthermore, the GA's evaluation cost could be reduced by incorporating zero-shot or learning-curve extrapolation proxies~\cite{maleki2026qnasneuralarchitecturesearch}, accelerating the search across larger populations and longer generation budgets. Finally, scaling to multiclass tasks beyond MNIST and evaluating on domain-specific benchmarks such as quantum chemistry or medical imaging would broaden the evidence base for Q-PhotoNAS as a general-purpose design tool.

 \section{Conclusion}

We presented Q-PhotoNAS, the first neural architecture search framework specifically designed for hybrid photonic quantum-classical architectures, by jointly searching classical and quantum design choices via genetic evolution over $3.7 \times 10^{10}$ configurations that is intractable to manual tuning. Learnable phase encoding enables the photonic circuit to align its phase space with the data distribution end-to-end. Evaluated on two image classification benchmarks, Q-PhotoNAS achieves final validation accuracies of 99.44\% on Digits and 98.78\% on MNIST, with the GA-found architectures outperforming manually designed baselines by a substantial margin. Quantum contribution analysis confirms that the photonic layer produces Fock-basis representations with low inter-class cosine similarity (0.135 for Digits, 0.100 for MNIST) and distinct quantum-classical feature relationships (per-sample cosine similarity of 0.167 for Digits and $-$0.154 for MNIST), demonstrating that the photonic component extracts information complementary to the classical pathway rather than replicating it; the negative overlap on MNIST indicates active anti-alignment, a stronger form of complementarity. Hybrid models consistently outperform matched-parameter classical baselines by approximately 0.65\,pp on Digits and 0.83\,pp on MNIST across three independent seeds, with lower variance than the classical baseline, providing an advantage within the hybrid pipeline. Hardware execution time analysis on the Quandela Ascella thermal QPU confirms single-image inference of $67.09 \pm 2.55$\,ms (Digits) and $148.61 \pm 5.78$\,ms (MNIST), with the photonic quantum component dominating latency in both cases ($>$98\%). Thermal phase-shifter reconfiguration is identified as the primary bottleneck, scaling linearly with circuit depth, and represents the highest-leverage target for future hardware optimization. Combined, these results establish that automated architecture search is both practical and impactful for hybrid photonic quantum systems. Q-PhotoNAS opens a principled path toward systematic design of classical and photonic components on quantum hardware, and provides a reusable framework that can be extended to larger mode counts, alternative encoding strategies, and real hardware deployment.


\bibliographystyle{IEEEtran}

\bibliography{refs}

\begin{IEEEbiography}[{\includegraphics[width=1in,height=1.25in,clip,keepaspectratio]{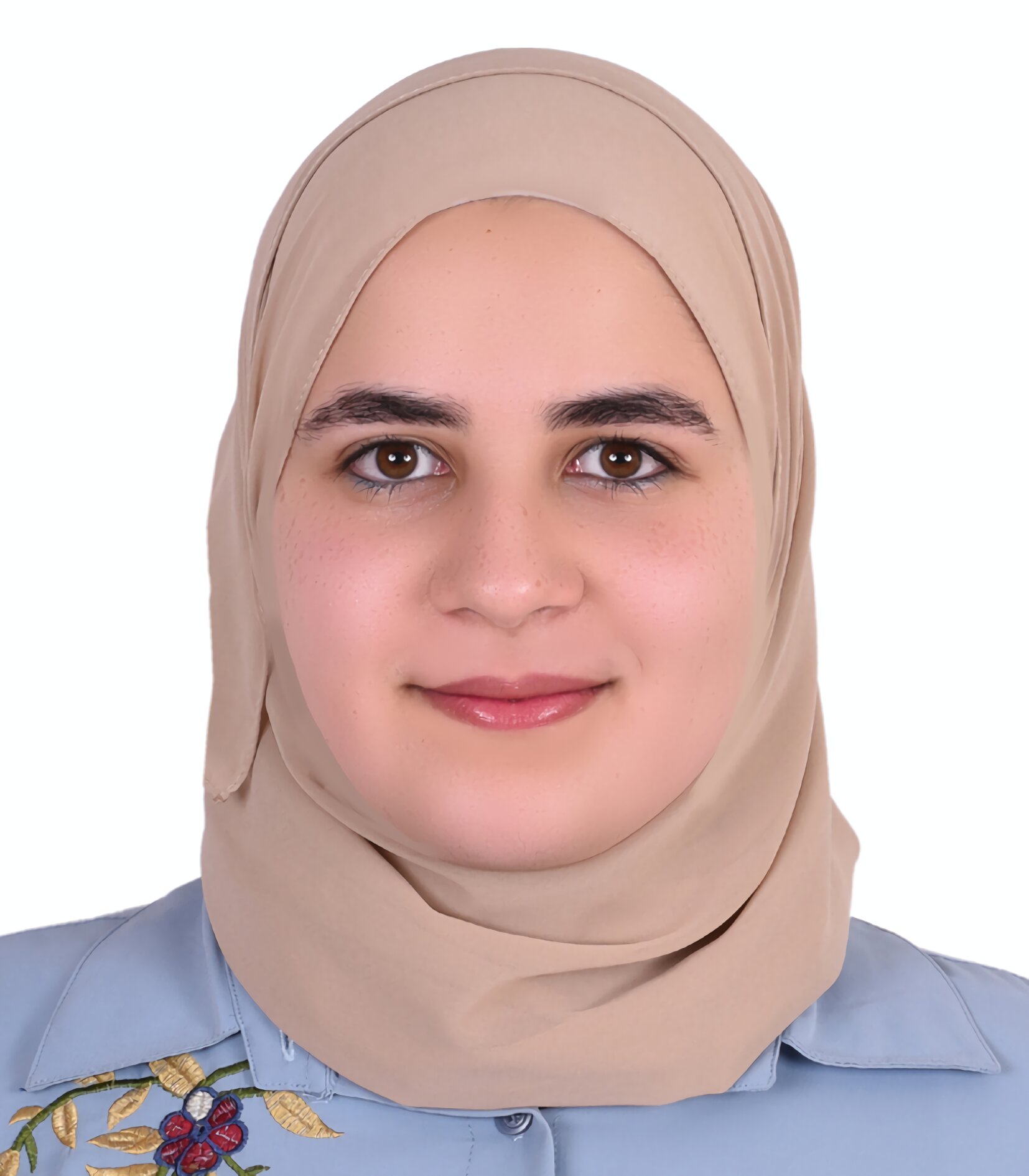}}]{Farah Elnakhal} is an undergraduate student in Computer Science and Mathematics at New York University Abu Dhabi (NYUAD), United Arab Emirates. She is a member of the eBrain Lab, where her research focuses on photonic quantum computing, with an emphasis on optimization techniques and noise for near-term quantum systems.
\end{IEEEbiography}

\begin{IEEEbiography}[{\includegraphics[width=1in,height=1.25in,clip,keepaspectratio]{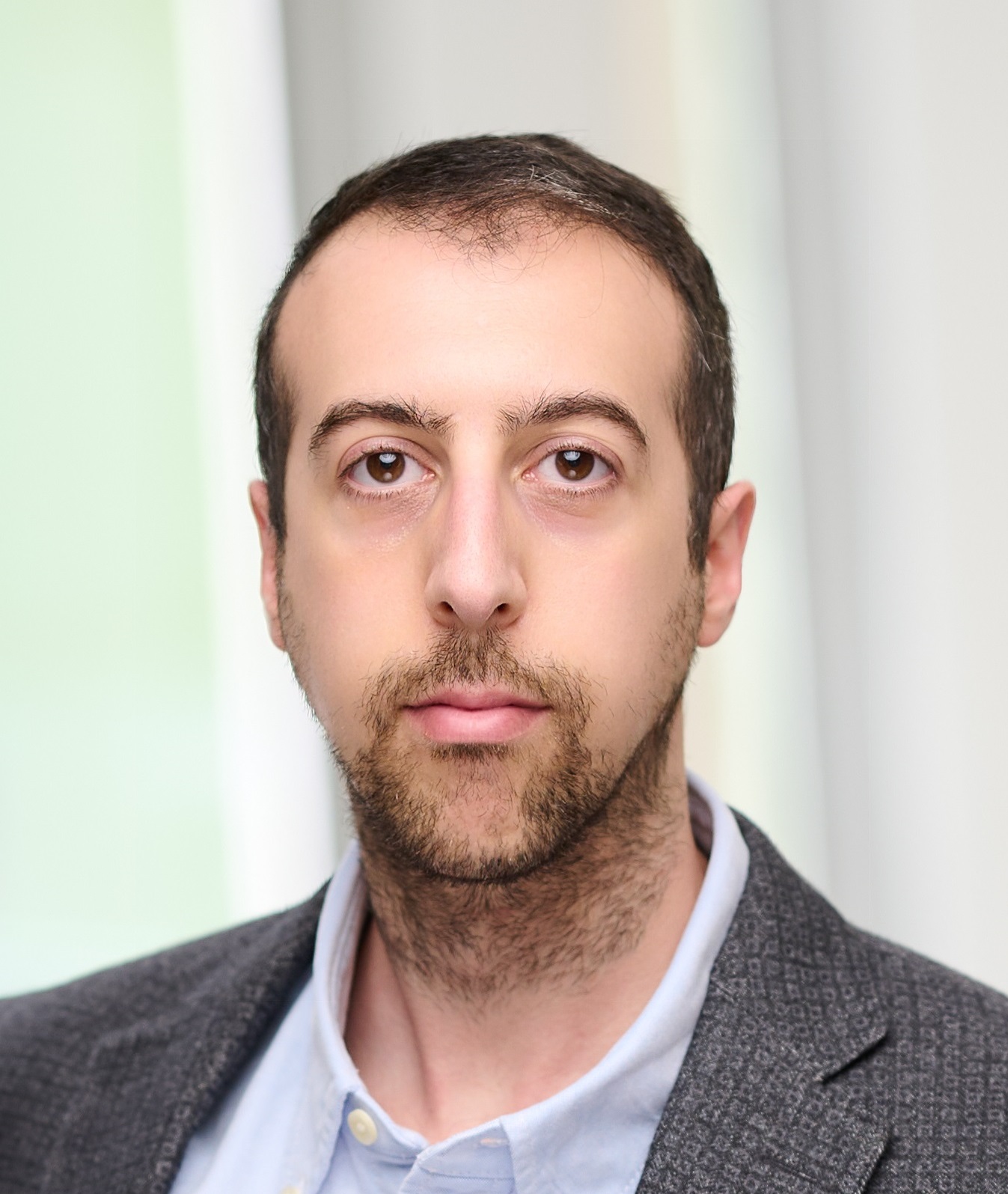}}]{Alberto Marchisio} (Member, IEEE) received
the B.Sc. and M.Sc. degrees in electronic engineering from Politecnico di Torino, Turin, Italy,
in October 2015 and April 2018, respectively,
and the Ph.D. degree in computer science
from Technische Universität Wien (TU Wien),
Vienna, Austria, in September 2023. Currently,
he is a Research Group Leader with the eBrain
Lab, Division of Engineering, New York
University Abu Dhabi (NYUAD), United Arab
Emirates. He has co-authored more than 70 papers in international
conferences/journals. His research interests include hardware and software
optimizations for machine learning and quantum machine learning,
EdgeAI, GenAI, neuromorphic computing, robust design, and approximate
computing for energy efficiency. He received several excellence awards,
including the Richard Newton Young Fellow Award, in 2019, and
the 2023 International Neural Network Society Doctoral Dissertation Award
runner-up at the World Congress of Computational Intelligence (WCCI
2024).
\end{IEEEbiography} 

\begin{IEEEbiography}[{\includegraphics[width=1in,height=1.25in,clip,keepaspectratio]{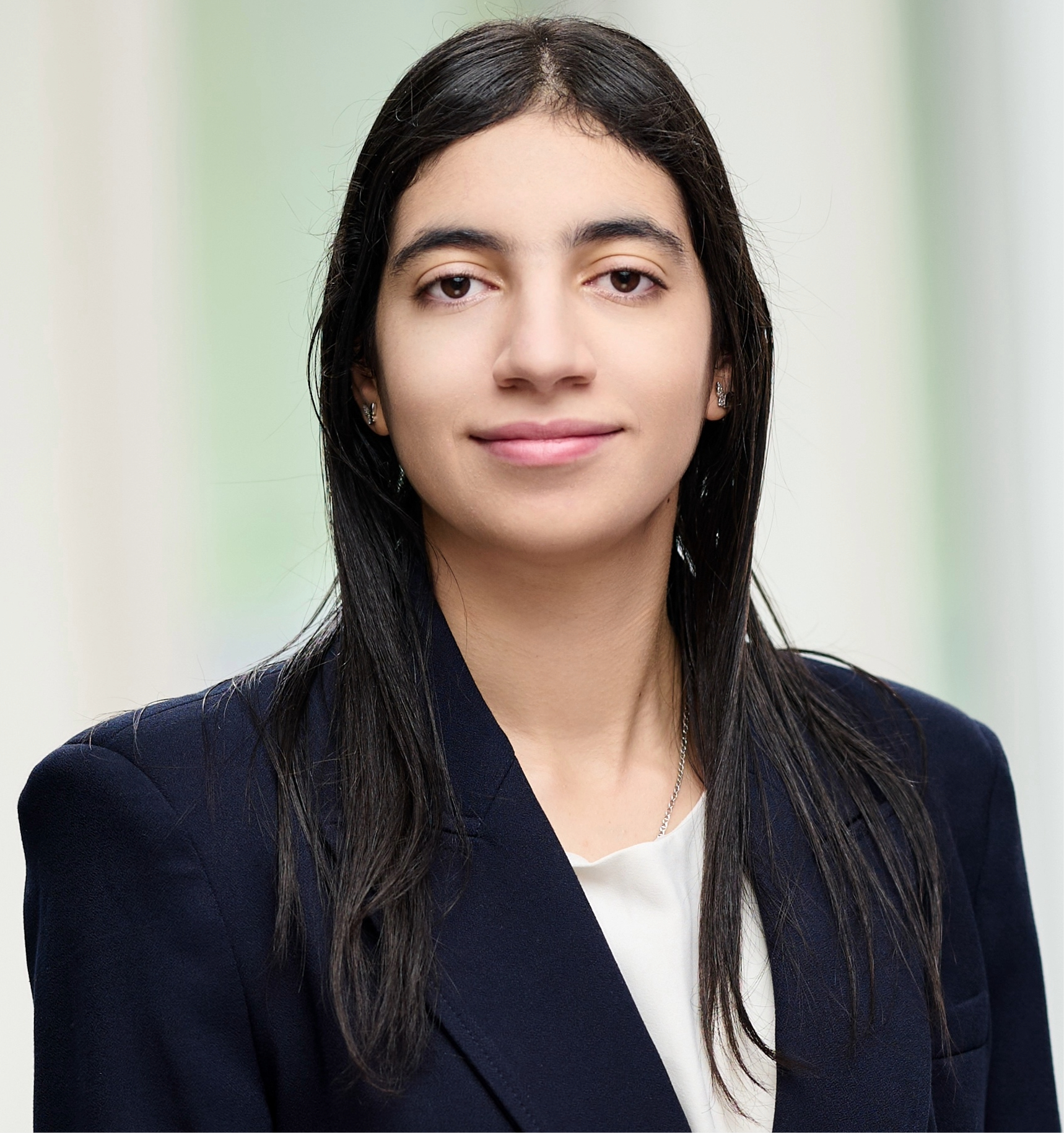}}]{Nouhaila Innan} (Member, IEEE) received her Ph.D. in Quantum Machine Learning from Hassan II University of Casablanca. She is currently a Postdoctoral Associate at the Center for Quantum and Topological Systems, New York University Abu Dhabi, and a Research Team Lead at eBRAIN Lab. Her research lies at the intersection of quantum computing, machine learning, and intelligent systems, with a focus on quantum machine learning, quantum neural networks, quantum federated learning, quantum optimization, and hybrid quantum-classical algorithms. Her work develops and evaluates quantum-enhanced learning models for applications in cybersecurity, finance, healthcare, network intelligence, and autonomous systems, with emphasis on privacy-preserving learning, hardware-aware design, and benchmarking for near-term quantum systems. She has served as a program committee member and reviewer for several international conferences and journals in quantum computing, artificial intelligence, and machine learning. She is also involved in leading research projects, mentoring students, organizing scientific initiatives, and supporting the growth of quantum computing education and research regionally and internationally.
\end{IEEEbiography}

\begin{IEEEbiography}[{\includegraphics[width=1in,height=1.25in,clip,keepaspectratio]{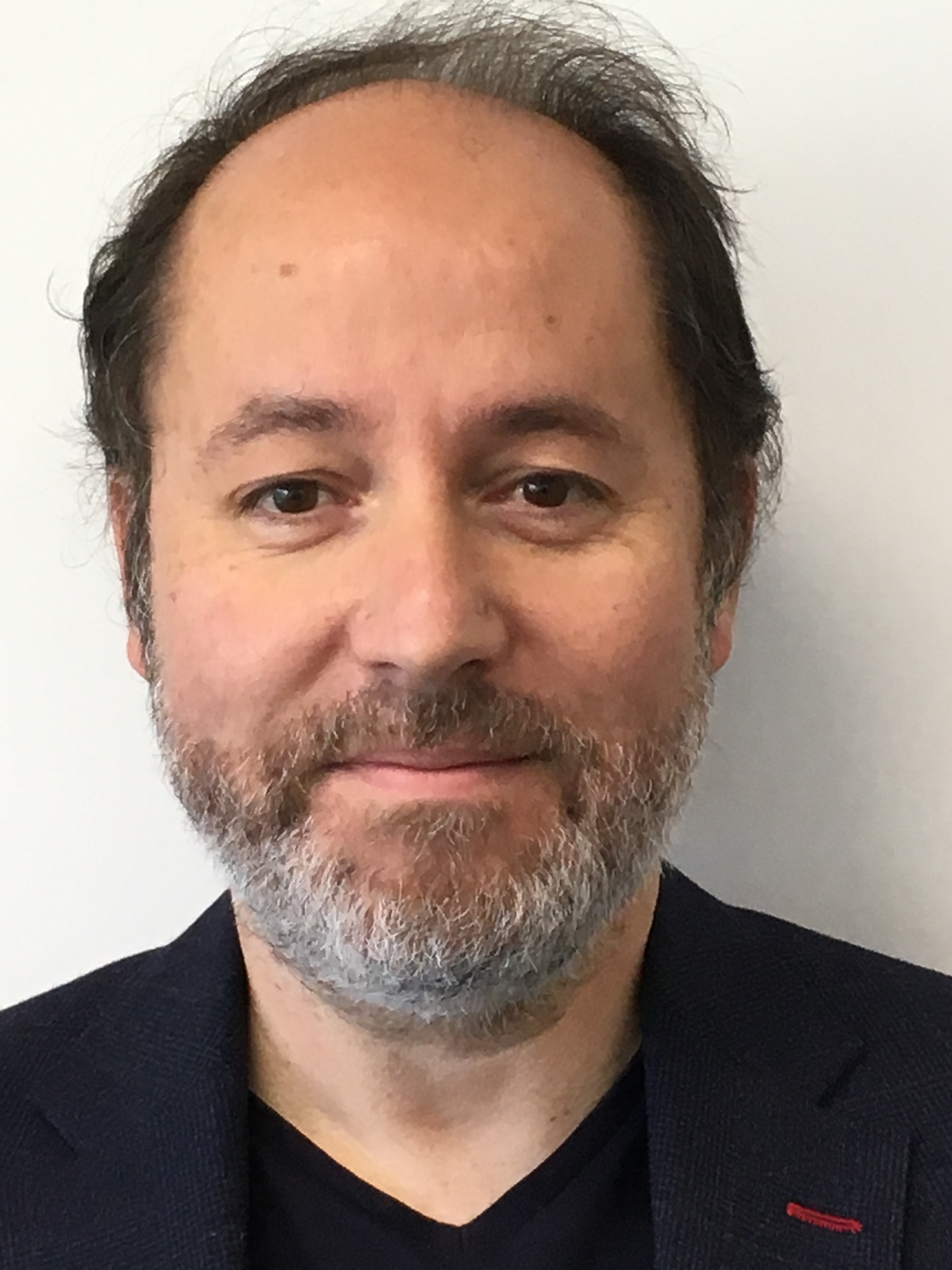}}]{Gabriel Falcao} 
(S'07--M'11--SM'14, Senior Member IEEE) graduated in Electrical and Computer Engineering (ECE) from the University of Porto and received the Ph.D. degree from the University of Coimbra in 2010 in parallel computer architectures applied to communications problems (error correction and LDPC codes). He is a Tenured Associate Professor at the Department of Electrical and Computer Engineering of the Faculty of Science and Technology of the University of Coimbra, and a Senior researcher at Instituto de Telecomunicações (IT). In 2011/12 he was a PostDoc and again in 2017/18 a Visiting Professor at EPFL, Switzerland, where he developed work in the area of parallel computer architectures trying to bridge the worlds of CPU, GPU and FPGA programming. In 2018 he was a Visiting Professor at ETHZ, Switzerland, where he developed pioneering work on processing-in-memory systems. His research interests include the development of novel quantum and parallel computing paradigms to deal with compute-intensive signal processing applications, in particular those related with data communications and medical imaging. Gabriel Falcao was the recipient of a Google Faculty Research Award (together with J. Barreto) from Google Inc. in 2014 and he is the PI of a CUDA Research Center sponsored by NVIDIA to the Multimedia Signal Processing Lab, a R\&D unit in IT hosted at the Department of ECE of the University of Coimbra. He participated in several projects funded by the European Commission, including SpaceRider and EuroHPC, by FCT (as PI and co-PI), and by P2020 in partnership with industry. In 2020 he was General Co-Chair of IEEE SiPS and Local Chair of Euro-Par. He successfully supervised more than 40 MSc students (all completed), 6 PhD students (all completed) + 4 ongoing, and 2 PostDocs, published over 150 papers in international journals and conferences, and holds an international patent (Europe, USA, Japan) commercially explored by the medical industry. He is a member of the HiPEAC Network of Excellence, ACM SIGMICRO, IEEE Signal Processing Society, and an IEEE Senior Member. Gabriel is Associate Editor of the IEEE Transactions on Signal Processing and IEEE Micro. He co-edited with John L. Hennessy and Christos Kozyrakis the Special Issue on “The Past, Present, and Future of Warehouse-Scale Computing” at IEEE Micro Magazine, in Sep/Oct 2024.
\end{IEEEbiography}

\begin{IEEEbiography}[{\includegraphics[width=1in,height=1.25in,clip,keepaspectratio]{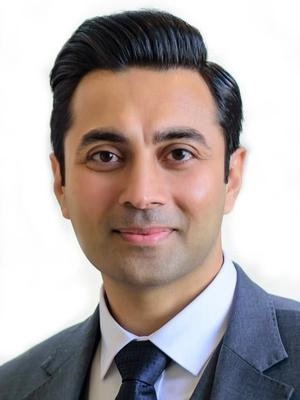}}]{Muhammad Shafique (M'11 - SM'16)} (Senior Member, IEEE)  received the Ph.D. degree in computer science from the Karlsruhe Institute of Technology (KIT), Germany, in 2011. In Oct.2016, he joined the Faculty of Informatics at TU Wien, Vienna, Austria as a Full Professor of Computer Architecture and Robust, Energy-Efficient Technologies. Since Sep.2020, Dr. Shafique is with the New York University (NYU), where he is currently a Full Professor and the director of eBRAIN Lab and iCAS Lab at the NYU-Abu Dhabi in UAE, and a Global Network Professor at the Tandon School of Engineering, NYU-New York City in USA. He is also a Co-PI/Investigator in multiple NYUAD Centers on Cybersecurity, Quantum Computing, AI \& Robotics, and Smart Cities. His research interests are in AI \& machine learning hardware and system-level design, brain-inspired computing, EdgeAI, tinyML, machine learning security and privacy, quantum machine learning, cognitive autonomous systems, wearable healthcare, AI for healthcare/medical imaging, energy-efficient systems, robust computing, hardware security, emerging technologies, electronic design automation, FPGAs, MPSoCs, and embedded systems. The researched technologies and tools are deployed in application use cases from IoT, Smart CPS, Healthcare and Robotics domains.
Dr. Shafique has given several Keynotes, Invited Talks, and Tutorials, as well as organized many special sessions at premier venues. He has served as the PC Chair, General Chair, Track Chair, and PC member for several prestigious IEEE/ACM conferences. Dr. Shafique holds one U.S. patent, and has (co-)authored 10 Books, 25+ Book Chapters, 450+ papers in premier journals and conferences, and 200+ archive articles. He received the 2015 ACM/SIGDA Outstanding New Faculty Award, the AI-2000 Chip Technology Most Influential Scholar Awards (2020, 2022, 2023; Honorable Mention 2024, 2025), the ASPIRE AARE Research Excellence Award in 2021, six gold medals, several best paper awards and nominations at prestigious conferences, several HiPEAC paper awards, and multiple competition awards. He is a senior member of the IEEE and IEEE Signal Processing Society (SPS), and a senior member of the ACM, SIGARCH, SIGDA, SIGBED, and HIPEAC.
\end{IEEEbiography}

\EOD
\end{document}